\documentclass[journal]{IEEEtran}
\usepackage{xcolor,soul,framed} 

\usepackage[ruled,linesnumbered]{algorithm2e}
\usepackage{tabularx}
\colorlet{shadecolor}{yellow}
\usepackage[pdftex]{graphicx}
\graphicspath{{../pdf/}{../jpeg/}}
\DeclareGraphicsExtensions{.pdf,.jpeg,.png}

\usepackage[cmex10]{amsmath}
\usepackage{array}
\usepackage{mdwmath}
\usepackage{mdwtab}
\usepackage{eqparbox}
\usepackage{url}
\usepackage{booktabs}
\usepackage{multirow}
\usepackage{caption}
\usepackage[normalem]{ulem}

\usepackage{subcaption}

\begin{document}

\title{Emergent Communication for Co-constructed Emotion Between Embodied Agents via \\Collective Predictive Coding}

\author{Zehang~Zhang,~\IEEEmembership{Student~Member,~IEEE,}
        Nguyen~Le~Hoang,~\IEEEmembership{Member,~IEEE,}
        Tadahiro~Taniguchi,~\IEEEmembership{Member,~IEEE,}
        and~Takato~Horii,~\IEEEmembership{Member,~IEEE,}

\thanks{This work was supported by JSPS KAKENHI Grant Number JP23H04834 and 23H04835.}
\thanks{Z. Zhang is with the Graduate School of Engineering Science, The University of Osaka, Toyonaka 560-8531, Japan.}%
\thanks{N. L. Hoang and T. Taniguchi are with the Graduate School of Informatics, Kyoto University, Kyoto 606-8501, Japan.}%
\thanks{T. Horii is with the Graduate School of Engineering Science, The University of Osaka, Toyonaka 560-8531, Japan. (e-mail: takato@sys.es.osaka-u.ac.jp)}%
\thanks{Manuscript received Month DD, YYYY; revised Month DD, YYYY.}}

\markboth{IEEE Transactions on Cognitive and Developmental Systems}%
{Zhang \MakeLowercase{\textit{et al.}}: Co-construction of Emotion Based on CPC}

\maketitle

\begin{abstract}
According to the theory of constructed emotion, the brain actively forms emotion categories by integrating multimodal bodily signals, and constructs emotional experiences by using these categories to predict and interpret sensory inputs. While research has advanced in modeling individual emotion construction, the social process of co-construction—how a shared understanding of emotions emerges between individuals—remains computationally underexplored. This study investigates this process by modeling emergent communication between two embodied agents using the Metropolis-Hastings Naming Game (MHNG), grounded in the Collective Predictive Coding (CPC) framework. Our experiments, using visual, auditory, and simulated interoceptive inputs, yield two main findings. First, MHNG-based communication significantly improves the alignment, clarity, and inter-agent agreement of the learned emotion categories compared to non-communicative and non-selective baselines, with the alignment effect concentrated at the symbolic layer rather than the perceptual latent representation. Second, even when the two agents have systematically divergent interoceptive dynamics, communication still produces robust categorical alignment, with distinct, category-specific reshaping patterns of each agent's emotion categories—consistent with the constructed-emotion view that interoceptive heterogeneity is constitutive of, rather than an obstacle to, shared emotional meaning. These findings provide computational support for the co-constructionist view of emotion and extend the CPC framework from physical to socially-grounded domains.
\end{abstract}

\begin{IEEEkeywords}
Theory of constructed emotion, predictive coding, symbol emergence, emergent communication, Metropolis-Hastings, naming game, deep generative model, machine learning.
\end{IEEEkeywords}

\IEEEpeerreviewmaketitle
\section{Introduction}
\begin{figure*}[t]
  \begin{center}
    \includegraphics[width=\linewidth]{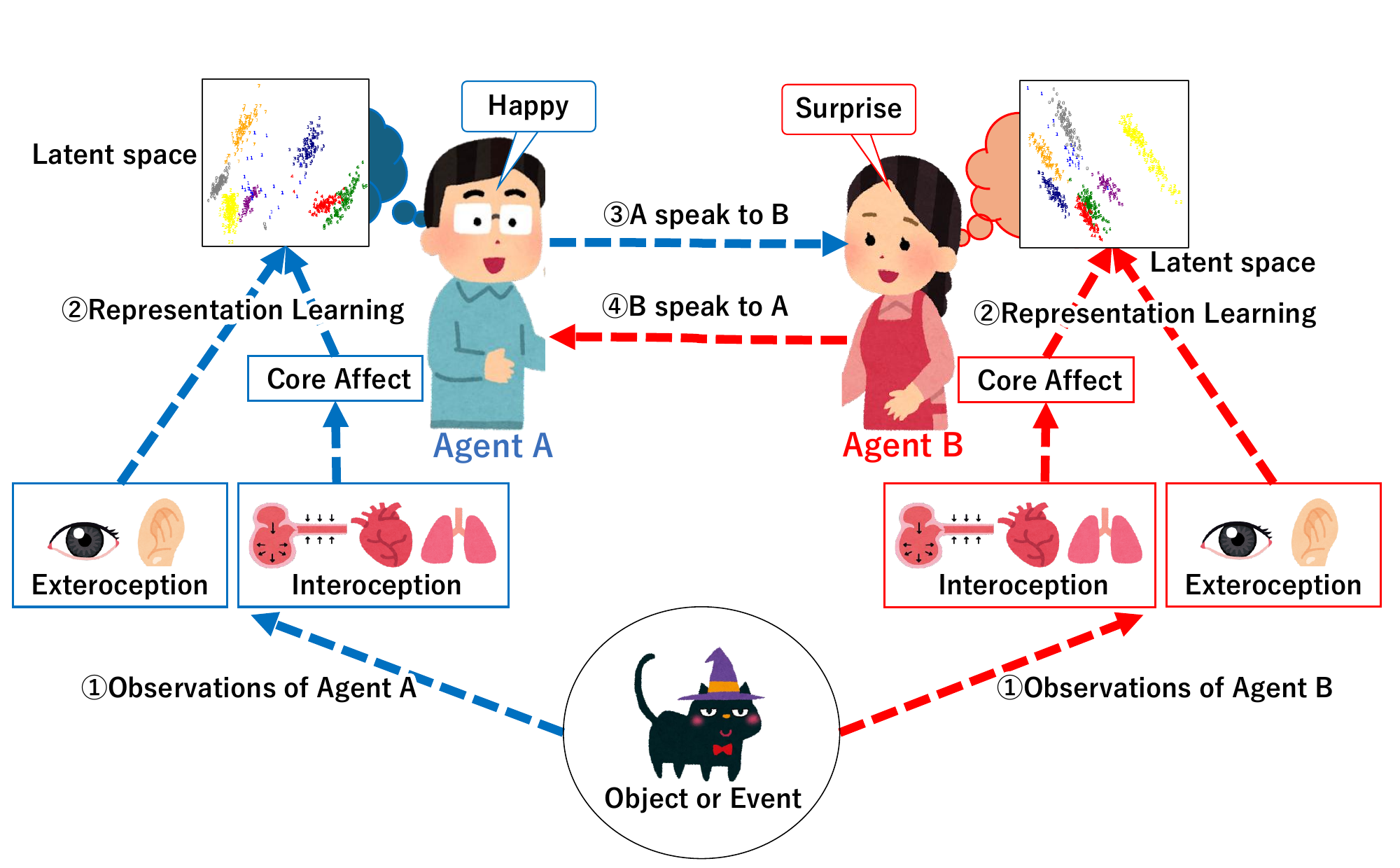}
    \caption{The process of two agents (Agent A and Agent B) forming and sharing emotion categories after observing the same object through communication. Each agent receives the similar external stimulus (A joint attention object) and responds to it through exteroception and interoception (e.g., heartbeat, visceral state, etc.).Interoception is integrated within the agent to generate core affect, which is further integrated with exteroception to promote the formation of emotion categories. Subsequently, the agents communicate symbolically (language interaction) based on emotion category inference. Through this emotion communication mechanism, the two agents dynamically update their respective emotion categories and achieve the co-construction of emotion.}
    \label{fig:emotion Communication}
  \end{center}
  \vspace{-1mm}
\end{figure*}

\IEEEPARstart{H}{ow} do we acquire concepts of emotions from our own sensory experiences? And how do we come to be able to share and understand those emotions with others?

Emotions are understood as internal psychological states that profoundly affect individual cognition, experience, and behavior, constituting complex responses to environmental stimuli \cite{Lazarus1991EmotionAndAdaptation}.
Of particular note in the multifaceted nature of emotions is their physical embodiment: emotions encompass not only subjective feelings but also physiological reactions, cognitive evaluations, and behavioral tendencies \cite{Scherer2005WhatAreEmotions}.
Recent findings emphasize the role of interoception—sensory signals originating within the body—as a crucial determinant in the experience and formation of emotions \cite{Damasio1999TheFeelingOfWhatHappens}.
These interoceptions are thought to eventually become integrated into a basic state known as core affect, which represents feelings of valence (pleasure--displeasure) and arousal (high--low activation)~\cite{russell1980circumplex}. Unlike specific emotion categories, core affect acts as the continuous physiological basis from which emotions are constructed.

Beyond internal mechanisms, emotions are also modulated by external social and cultural contexts — cross-cultural studies have demonstrated that the expression, perception, and regulation of emotions are significantly influenced by cultural frameworks \cite{Mesquita1992CulturalVariationsInEmotions, Kitayama1994EmotionAndCulture}. Yet despite emotion's critical role in cognition and social life, it remains insufficiently understood how emotion categories are constructed and how they come to be shared among individuals~\cite{barrett2017emotions,gendron2018emotion,Damasio1994DescartesError}.

Among theories of emotion, Ekman's basic emotion theory posits that a small set of emotions—such as joy, sadness, anger, fear, disgust, and surprise—are biologically hardwired, each underpinned by a distinct neural circuit, and universally expressed across cultures \cite{Ekman1992AnArgumentForBasicEmotions}. However, this essentialist view has faced substantial criticism. Meta-analyses of neuroimaging studies have failed to identify consistent, emotion-specific brain signatures \cite{lindquist2012brain,barrett2006emotions}, and cross-cultural research has revealed significant variation in how emotions are recognized and categorized \cite{russell1994there,gendron2014perceptions}, challenging the assumption of universality.

In contrast, the theory of constructed emotion \cite{barrett2017emotions} challenges this essentialist perspective. It argues that humans do not possess biologically hardwired reactions for specific emotions. Instead, the theory distinguishes between emotion categories and emotional experiences. Emotion categories are conceptual knowledge structures formed through the integration of interoception and exteroception, shaped by personal beliefs and memories. Emotional experiences, in turn, are the specific instances constructed when the brain applies these categories to predict and interpret continuous sensory inputs. In Barrett's framework, the formation of emotion categories is a prerequisite for these experiences; without the corresponding category, an individual would perceive only raw bodily sensations rather than specific emotions.
Specifically, this construction unfolds in two conceptual stages. First, the brain acquires emotion categories (e.g., "anger", "joy") learned from statistical regularities in past interoceptive and exteroceptive experiences. Second, specific emotion are experienced by applying these categories to current sensory inputs to resolve ambiguity. For instance, an accelerated heartbeat (interoception) is merely physiological noise until the brain categorizes it. If the brain predicts this state as "anxiety" in a stressful context, it is experienced as distress; if categorized as "excitement" in a positive context, the same physiological state is perceived as thrill. Thus, the emotion category shapes the subjective quality of the experience.
This constructionist view begs for a computational explanation of how such category formation and predictive experience might be implemented in the brain. 

The theory of predictive coding, proposed by Friston \cite{Friston:2017}, offers a unifying computational framework that provides a natural explanation for these processes. In this framework, the brain continuously constructs internal models that predict incoming sensory stimuli and strives to minimize the discrepancy between predictions and actual inputs, a discrepancy termed prediction error. Importantly, predictive coding extends beyond exteroceptive signals to interoceptive signals, as argued by Seth et al.~\cite{seth2016active}, who proposed that emotional experiences arise from minimizing prediction errors related to internal bodily states. Thus, predictive coding serves as a theoretical foundation that complements and reinforces the mechanisms proposed by the theory of constructed emotion, framing emotional experiences as emergent outcomes of hierarchical predictive processes.

Building on this approach, prior research has explored the internal formation of emotion categories based on embodied sensory inputs \cite{Horii:2018, hieida2018deepemotioncomputationalmodel}, but did not address how shared structures of emotion categories could emerge between individuals—a key aspect in understanding the social and cultural co-construction of emotion.
In contrast, Gendron and Barrett~\cite{gendron2018emotion} proposed a framework in which emotion perception is a dynamic process of \emph{conceptual synchrony} between two individuals, mediated in part by emotion words. While theoretically influential, their proposal has not been instantiated computationally.

Addressing this gap requires a computational framework capable of explaining how shared meaning emerges between individuals through interaction. The theory of symbol emergence \cite{taniguchi2016symbol} provides such a foundation: shared symbols are not predefined but arise through a bottom-up process in which agents form internal categories from their own sensory experiences and gradually align these categories through communicative interaction \cite{hagiwara2019symbol}. Taniguchi et al. further formalized this process through CPC \cite{taniguchi2024cpc}, which extends predictive coding from the individual to the social level — agents not only minimize prediction errors within their own sensory models but also coordinate their internal representations through the exchange of symbolic signs, thereby achieving inter-agent alignment. The co-construction of emotion maps naturally onto this framework: emotion categories, formed from each individual's unique bodily and sensory experience, can be viewed as internal representations that become socially aligned through communicative exchange. This study thus situates the co-construction of emotion within the CPC framework and models it as a process of CPC, as conceptually illustrated in Fig.~\ref{fig:emotion Communication}.

Specifically, we employ the Inter-Gaussian Mixture Model+Multivariate Variational Autoencoder (Inter-GMM+MVAE) \cite{hoang10emergent}, which integrates a multimodal deep generative model with the MHNG. We use this framework to simulate emergent communication—a bottom-up process where shared symbol systems evolve through local interactions without external supervision—between two agents processing visual, auditory, and interoceptive inputs. Our primary goal is to investigate how this communication influences the formation and alignment of emotion categories. To do so, we compare the structure and coherence of categories acquired in scenarios with and without MHNG-based communication. 
Crucially, the theory of constructed emotion posits that physiological variations are ubiquitous across individuals; the same category (e.g., "anger") may be constructed from different interoception depending on the person \cite{barrett2017emotions}. This raises a fundamental question: does shared emotional understanding require agents to have identical bodies, or can it be achieved despite physiological differences? Gendron and Barrett \cite{gendron2018emotion} suggest that emotion perception relies on "conceptual synchrony" rather than physiological mirroring. To computationally validate this hypothesis, we conduct experiments introducing agents with divergent interoceptive dynamics. By manipulating the similarity of interoception in the two agents, we investigate whether emergent communication can bridge the "interoceptive gap" and enable the co-construction of shared emotion categories even in the absence of physiological isomorphism.
To probe the crucial role of embodied reaction, we conduct additional experiments manipulating the agents' interoceptive inputs. We specifically examine conditions where embodiment---specifically interoceptive reactivity---differs between the two agents, in addition to conditions with a complete absence of interoception.

The main contributions of this study are as follows:
\begin{itemize}
    \item We apply Inter-GMM+MVAE, a computational model of CPC, to model the co-construction of emotion and empirically confirm the effect of emergent communication on emotional alignment.
    \item We investigate the effect of interoceptive differences on the co-construction of emotion by introducing agents with systematically varied embodiments. Our experiments demonstrate that MHNG-based communication not only supports the formation of similar emotion category structures between agents, but also absorbs differences in bodily reactions, enabling robust co-construction of emotion even when the two agents have divergent interoceptive dynamics.
\end{itemize}
\section{Related Study}
    \subsection{Emotion model}
Several studies have explored how emotion categories emerge from embodied sensory experience, spanning both theoretical frameworks and computational models. Horii et al. \cite{Horii:2018} proposed a developmental model of emotion perception in which an infant agent learns to recognize a caregiver's emotions by integrating visual, auditory, and tactile signals through hierarchically structured restricted Boltzmann machines, showing that tactile dominance and perceptual improvement jointly facilitate emotion differentiation. Hieida et al. \cite{hieida2018deepemotioncomputationalmodel} developed a computational model of emotion that integrates visual stimuli with simulated interoceptive signals, framing emotion formation as a generative process driven by the interaction between internal and external appraisal. Seth \cite{seth2013interoceptive} proposed the interoceptive inference model, which applies predictive coding to interoception, arguing that emotional experiences arise from the brain's active inference on the causes of internal bodily signals rather than from passive bottom-up processing.
While these studies successfully examined the internal formation of emotion categories within a single agent, none addressed how shared emotion categories could emerge between individuals — the social co-construction of emotion process that is the focus of this study.

\subsection{Co-construction of Concepts}
Gendron and Barrett \cite{gendron2018emotion} proposed that emotion perception is not the passive detection of discrete signals but a dynamic process of conceptual synchrony between two individuals. Drawing on predictive coding and grounded cognition, they argued that both the perceiver and the target continuously generate and refine predictions about each other's internal states through multimodal sensory cues such as facial expressions, vocal changes, and bodily movements. Language plays a central role in this process: emotion words serve as efficient activators of conceptual knowledge and as explicit bids for mutual understanding, enabling what they termed the co-construction of emotion. However, their framework remained at the theoretical level, without proposing specific computational models.

To computationally model such co-construction processes, the theory of symbol emergence systems (SES) \cite{taniguchi2016symbol} and its formalization through CPC \cite{taniguchi2024cpc} provide a promising foundation. In SESs, agents form internal representations through physical interactions with the environment and simultaneously organize shared external representations (symbols) through semiotic communication with other agents. Taniguchi formalized this dynamics through the CPC hypothesis, which extends predictive coding from a single brain to a multi-agent system: symbol emergence is cast as decentralized Bayesian inference, where agents collectively infer shared representations that maximize the predictability of their distributed sensory observations. The MHNG \cite{hagiwara2019symbol, taniguchi2023emergent} provides a concrete algorithmic realization of this process — agents exchange signs without explicit feedback, and the acceptance or rejection of proposed signs follows a Metropolis-Hastings criterion, guaranteeing convergence to the posterior distribution over shared symbols.

Several computational models have instantiated this framework with increasing complexity. Taniguchi et al. \cite{taniguchi2023emergent} proposed the Inter-GMM+VAE, the first deep generative model for emergent communication based on the MH naming game, in which two agents observing the same objects from different viewpoints cooperatively form internal representations via VAEs, learn categories via GMMs, and share signs without explicit feedback. Hagiwara et al. \cite{hagiwara2022multiagent} proposed the Inter-MDM, which extended the framework to multimodal agents equipped with visual, auditory, and haptic modalities, demonstrating that emergent communication enables agents to form shared categories even when some sensory modalities are missing. Hoang et al. \cite{hoang10emergent} further advanced the approach with the Inter-GMM+MVAE, integrating multimodal VAEs with the MH naming game and systematically comparing fusion strategies (PoE, MoE, MoPoE), showing that PoE consistently produced the most effective latent spaces for emergent communication. Beyond object categorization, Sakurai et al. \cite{sakurai2026mh} demonstrated the generality of the CPC framework by applying it to collaborative music generation through MH-MuG, where agents with distinct musical knowledge collectively produce stylistically fused compositions via decentralized Bayesian inference.
While these models have been applied to object categorization and music generation, no prior work has applied the CPC framework to the domain of emotion — the focus of the present study.

\section{Preliminaries}
We adopt the Inter-GMM+MVAE framework proposed by Hoang et al.~\cite{hoang10emergent}, which integrates MVAE with a MHNG to support symbol emergence between agents. In this section, we first introduce the Product-of-Experts MVAE (PoE-MVAE), which serves as the generative model for integrating interoceptive and exteroceptive information. Second, we describe the generation of core affect, which serves as the continuous physiological basis for the agent's emotional experience and acts as a crucial input to the model. Finally, we detail the MHNG, the mechanism that enables agents to communicate and align emotion categories without explicit feedback.

\subsection{Multimodal Variational Autoencoder(MVAE)}
The theory of constructed emotion posits that emotion categories are formed through the integration of multiple sensory streams — including interoception, visual perception, and auditory input — rather than from any single modality alone \cite{barrett2017emotions}. Computationally modeling this process therefore requires a generative framework capable of fusing heterogeneous sensory inputs into a unified latent representation. The MVAE provides such a framework: it learns a shared latent space from multiple modalities, enabling both the integration of complementary information and the reconstruction of individual modalities from this shared representation \cite{wu2018multimodal}.
In this study, we adopt the PoE-MVAE~\cite{wu2018multimodal} to integrate multiple modalities into a shared latent representation. PoE achieves this by taking the product of posterior distributions from modality-specific encoders, resulting in a consensus representation that emphasizes the overlapping beliefs across modalities. 

Formally, given modality-specific posterior approximations \( q_{\phi_m}(z \mid x_m) \) for each modality \( m \in \{1, \dots, M\} \), the joint latent posterior is defined as:
\begin{equation}
q_{\text{PoE}}(z \mid X) \propto \prod_{m=1}^{M} q_{\phi_m}(z \mid x_m),
\end{equation}
where \( z \) denotes the shared latent variable, \( x_m \) represents the observation for the \( m \)-th modality, and \( X = \{x_1, \dots, x_M\} \) is the set of all multimodal observations.This fusion mechanism makes PoE especially robust to noisy or missing modalities, while being able to generate a compact and consistent latent space. Compared with other multimodal fusion strategies, such as Mixture-of-Experts (MoE)~\cite{shi2019variational} and Mixture-of-Product-of-Experts (MoPoE)~\cite{sutter2021generalized}, PoE consistently demonstrates superior performance in both clustering metrics and latent space quality, particularly in emergent communication tasks using the MHNG~\cite{hoang10emergent}.

Critically for emotion modeling, PoE effectively resolves the ambiguity often present in individual sensory channels. For example, even if the posterior probability of an emotion category is ambiguous based on a single modality (e.g., a subtle facial expression), PoE integrates complementary signals from other modalities (e.g., voice or interoception) to produce a sharper, more confident joint posterior. In contrast, averaging-based methods like MoE tend to dilute such signals, resulting in flatter or multi-peaked distributions. This ability to yield a concentrated and unimodal latent representation not only handles multimodal ambiguity but also aligns closely with the Gaussian assumptions made by GMM, thereby facilitating more stable and effective parameter estimation and clustering performance.

\subsection{Metropolis-Hastings Naming Game (MHNG)}
The MHNG~\cite{hagiwara2019symbol, taniguchi2023emergent} is a probabilistic communication game that enables two agents to develop shared symbolic representations which does not rely on explicit feedback. In each interaction, both agents jointly attend to the same object, forming internal representations through their own sensations. One agent, designated as a speaker, samples a sign from a posterior distribution conditioned on its internal representation inferred from sensory signals and sends it to a listener. The listener evaluates the received sign based on its own inferred internal state, applying a MH acceptance criterion:
\begin{equation}
\label{equation:radio}
r = \min\left(1, \frac{P(z_{d}^{\mathrm{Li}}|\mu^{\mathrm{Li}}, \Lambda^{\mathrm{Li}}, w_{d}^{\mathrm{Sp}})}{P(z_{d}^{\mathrm{Li}}|\mu^{\mathrm{Li}}, \Lambda^{\mathrm{Li}}, w_{d}^{\mathrm{Li}})}\right),
\end{equation}
where $z_{d}^{\mathrm{Li}}$ is the listener’s latent representation, and $w_{d}^{\mathrm{Sp}}$ and $w_{d}^{\mathrm{Li}}$ denote the speaker's and listener’s signs, respectively. If the listener accepts the received sign according to this probability, it updates its internal model parameters accordingly. The roles of speaker and listener are then alternated, and the process is iteratively repeated across data points. 

The entire MHNG procedure is summarized in Algorithm~\ref{alg:mh_communication}. Algorithm~\ref{alg:mh_communication} details the computational steps of a single communicative interaction. First, the speaker generates a proposal sign \( w_d^{Sp} \) by sampling from its distribution conditioned on its latent state \( z_d^{Sp} \). Second, the listener evaluates this proposal against its own current sign \( w_d^{Li} \) by calculating the acceptance ratio \( r \) using equation~\ref{equation:radio}. This ratio effectively measures whether the speaker's sign explains the listener's internal state (\( z_d^{Li} \)) better than the listener's existing sign. Finally, the listener accepts the new sign \( w_d^{Sp} \) with probability \( \min(1, r) \), thereby dynamically aligning its symbolic representation with the speaker's without requiring direct access to the speaker's internal state.
\begin{algorithm}[t]
\caption{MHNG}
\label{alg:mh_communication}
\SetKwProg{Procedure}{\textbf{Procedure}}{}{end}

\Procedure{MHNG($z^{Sp}, \mu^{Sp}, \Lambda^{Sp}, z^{Li}, \mu^{Li}, \Lambda^{Li}, w^{Li}_d$)}{
    $w_d^{Sp} \sim P(w_d^{Sp} \mid z_d^{Sp}, \mu^{Sp}, \Lambda^{Sp})$\\
    $r \sim \min\left(1, \frac{P(z^{Li}_d \mid \mu^{Li}, \Lambda^{Li}, w_d^{Sp})}
                            {P(z^{Li}_d \mid \mu^{Li}, \Lambda^{Li}, w_d^{Li})}\right)$\\
    $u \sim \text{Unif}(0,1)$\\
    \eIf{$u \leq r$}{
        $w_d = w_d^{Sp}$
    }{
        $w_d = w_d^{Li}$
    }
}
\end{algorithm}

\subsection{Model of emotional communication:Inter-GMM+MVAE}
\begin{table*}[htbp]
\caption{Definition of variables used in the generative process of Inter-GMM+MVAE (Eqs.~(3)--(6)) and the inference algorithms (Algorithms~1 and~2). Variables with superscript~$*$ are agent-specific ($* \in \{A, B\}$)}
\centering
\begin{tabularx}{\textwidth}{lX}
\hline
\textbf{Symbol} & \textbf{Definition} \\
\hline
$* \in \{A, B\}$ & Agent identifier ($A$ and $B$). \\
$** \in \{i, v, a\}$ & Modality index: interoception ($i$), vision ($v$), and audio ($a$). \\
$K$ & Number of signs (categories). \\
$D$ & Number of data points. \\
$w_d$ & Discrete variable drawn from a categorical distribution; represents the assigned sign for the $d$-th data point. \\
$z_d^*$ & Latent variable inferred by each agent based on multimodal observations. \\
$o_{**,d}$ & Observed multimodal sensory data corresponding to each data point. \\
$\theta_{**}$ & Variational parameters associated with the VAE for each modality. \\
$\mu_k^*, \Lambda_k^*$ & Mean vector and precision matrix parameters of the kth multivariate normal distribution within the GMM. \\
$a, m, \beta, \nu, \pi$ & Hyperparameters for the distributions $\mu$, $\Lambda$, and $w$. \\

\hline
\end{tabularx}
\label{tab:variables}
\end{table*}
The Inter-GMM+MVAE~\cite{hoang10emergent} combines MVAE and MHNG to model emergent communication based on multimodal information. The two agents first learn the multimodal information of the joint-attention objects through MVAE to obtain the latent representation, and then modulate the sign learned by their respective GMMs through MHNG.

\begin{figure}[t]
  \begin{center}
    \includegraphics[width=\linewidth]{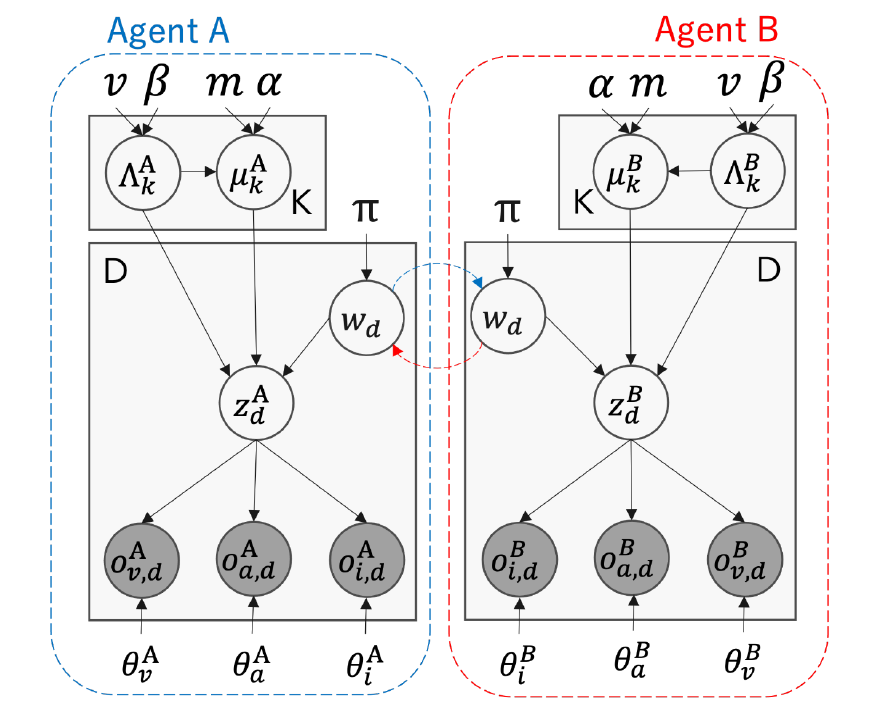}
    \caption{Graphical model of Inter-GMM+MVAE used in this research. Each agent $* \in \{A, B\}$ (left: blue, right: red) has its own GMM and MVAE.}
    \label{fig:Graphical model of Inter-GMM-MVAE}
  \end{center}
  \vspace{-1mm}
\end{figure}

The definition of Inter-GMM+MVAE can refer to Table~\ref{tab:variables} and Fig.~\ref{fig:Graphical model of Inter-GMM-MVAE}. Its generative process is defined as follows:

\begin{algorithm}[t]
\caption{Iterative Co-construction of Emotion Categories via Inter-GMM+MVAE}
\label{alg:inter_gmm_mvae}
\SetKwProg{}{}{}{end}

    Initialize all parameters\\
    \For{$t = 1$ \KwTo $T$}{
        \For{$d = 1$ \KwTo $D$}{
            $z^A_d \sim P(z^A_d \mid \theta^A, o^A_d, w_d^A, \mu^A, \Lambda^A)$\\
            $z^B_d \sim P(z^B_d \mid \theta^B, o^B_d, w_d^B, \mu^B, \Lambda^B)$\\
        }
        // Agent A speaks to Agent B\\
        \For{$d = 1$ \KwTo $D$}{
            $w_d^B \gets \text{MHNG}(z^A, \mu^A, \Lambda^A, z^B, \mu^B, \Lambda^B, w_d^B)$\\
        }
        // Learning by Agent B\\
        $\mu^B, \Lambda^B \sim P(\mu^B, \Lambda^B \mid w^B, z^B, \alpha^B, m^B, \beta^B, v^B)$\\
        $\theta^B \sim P(\theta^B \mid o^B, z^B)$\\
        // Agent B speaks to Agent A\\
        \For{$d = 1$ \KwTo $D$}{
            $w_d^A \gets \text{MHNG}(z^B, \mu^B, \Lambda^B, z^A, \mu^A, \Lambda^A, w_d^A)$\\
        }
        // Learning by Agent A\\
        $\mu^A, \Lambda^A \sim P(\mu^A, \Lambda^A \mid w^A, z^A, \alpha^A, m^A, \beta^A, v^A)$\\
        $\theta^A \sim P(\theta^A \mid o^A, z^A)$\\
    }

\end{algorithm}

\small
\begin{align}
    w_d &\sim \text{Cat}(\pi) & d &= 1, \ldots, D \label{gen-w} \\
    \mu_k^*, \Lambda_k^* &\sim \mathcal{N}(\mu_k^* | m, (\alpha\Lambda_k^*)^{-1})\mathcal{W}(\Lambda_k^* | \nu, \beta) & k &= 1, \ldots, K  \label{gen-mulambda} \\
    z_d^* &\sim \mathcal{N}(z_d^* | \mu_{w_d}^*, (\Lambda_{w_d}^*)^{-1}) & d &= 1, \ldots, D  \label{gen-z} \\
    o_{**, d}^* &\sim P_{\theta_{**}^*}(o_{**, d}^* | z_d^*) & d &= 1, \ldots, D \label{gen-o} 
\end{align}
Algorithm~\ref{alg:inter_gmm_mvae} outlines the iterative learning dynamics where perception, communication, and learning are interleaved. In each epoch, agents first infer their individual latent states \( z_d \) from multimodal observations. Subsequently, they engage in the naming game (MHNG) to update the shared signs \( w_d \). Crucially, these communicated signs then serve as pseudo-labels for the parameter update step: each agent optimizes its GMM parameters (\( \mu, \Lambda \)) and VAE decoder parameters (\( \theta \)) to maximize the likelihood of the observations given the agreed-upon signs. By alternating the roles of speaker and listener, both agents mutually adapt their internal representations to align with the socially constructed symbol system.
\normalsize

The interaction process between the two agents proceeds as follows:

\begin{enumerate}
\item[(1)] Both agents are exposed to the same emotional stimulus $E_d$ (e.g., watching a movie together) and each agent expresses individual reactions (such as facial expressions, vocalizations, and changes in interoception).
\item[(2)] Each agent observes their own reactions as observation data $o_{**, d}^*$ and infers the latent variable $z_d^*$ using their GMM+MVAE models.
\item[(3)] One agent takes the role of the speaker and generates a sign $w_d^{Sp}$ inferred from $z_d^{Sp}$, which is then sent to the another agent, i.e., the listener agent.
\item[(4)] (4) The listener agent computes the acceptance probability $r$ defined in Eq.~(2) and probabilistically accepts the speaker's sign. If accepted, the listener updates the parameters of its GMM+MVAE model according to the new sign; otherwise, it retains its previous sign.
\item[(5)] The roles of speaker and listener are then switched, and steps (3) and (4) are repeated.
\item[(6)] Steps (1) through (5) are iteratively repeated.
\end{enumerate}

\section{Experiment setting}
\begin{figure*}[htbp]
  \centering
  \begin{subfigure}[b]{0.24\textwidth}
    \includegraphics[width=\linewidth]{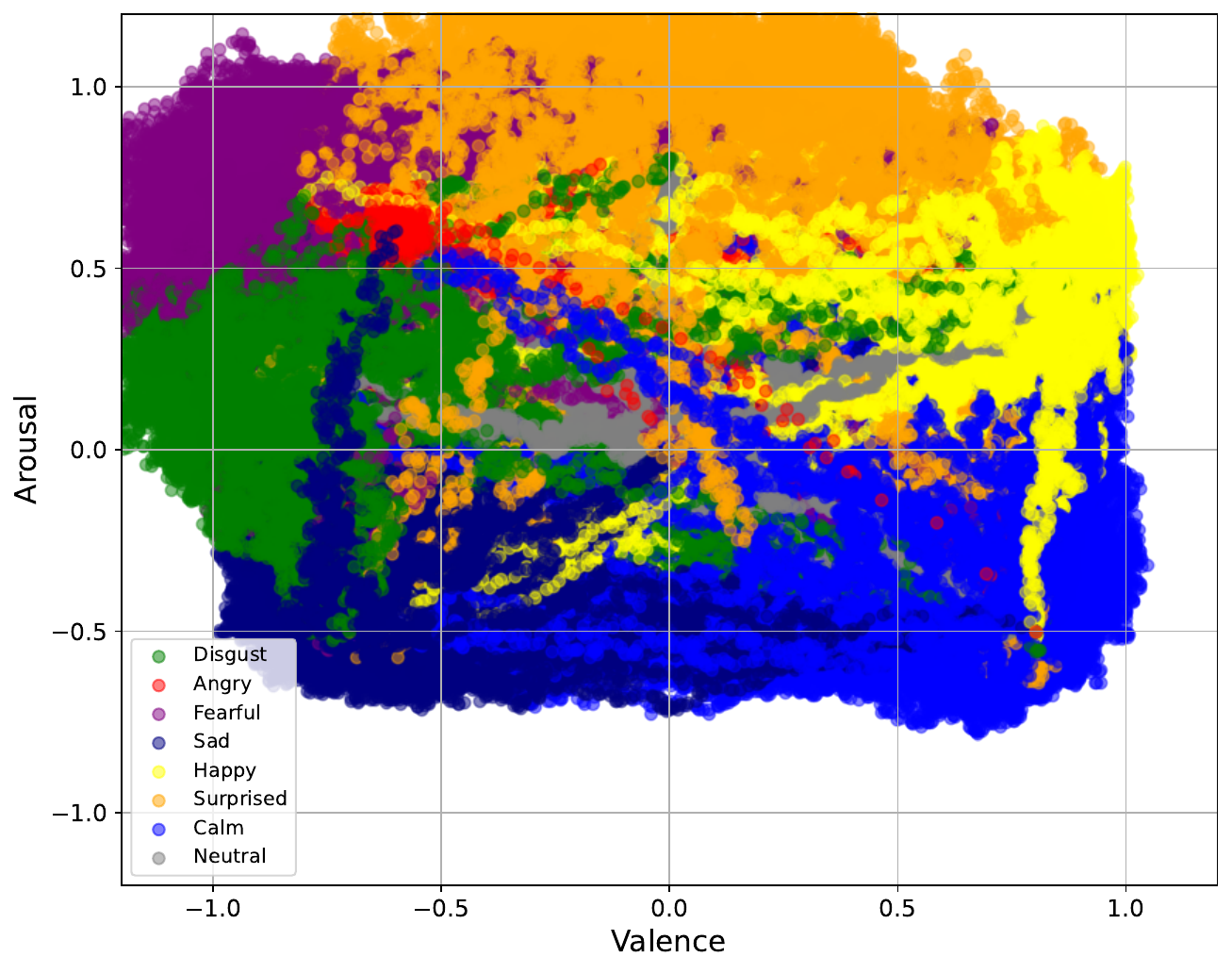}
    \caption{Original core affect}
    \label{fig:subfig1}
  \end{subfigure}
  \hfill
  \begin{subfigure}[b]{0.24\textwidth}
    \includegraphics[width=\linewidth]{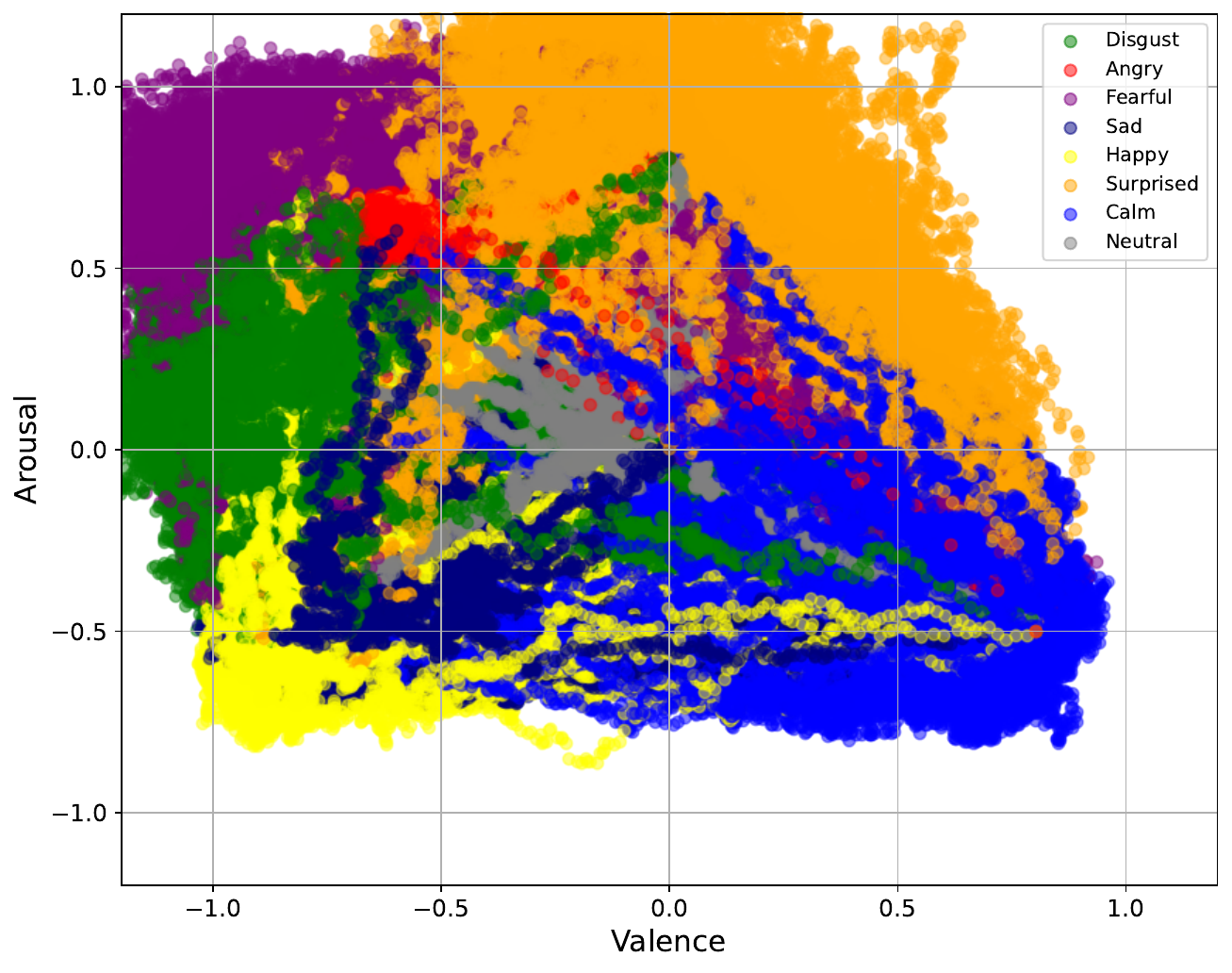}
    \caption{Happy inverse core affect}
    \label{fig:subfig2}
  \end{subfigure}
  \hfill
  \begin{subfigure}[b]{0.24\textwidth}
    \includegraphics[width=\linewidth]{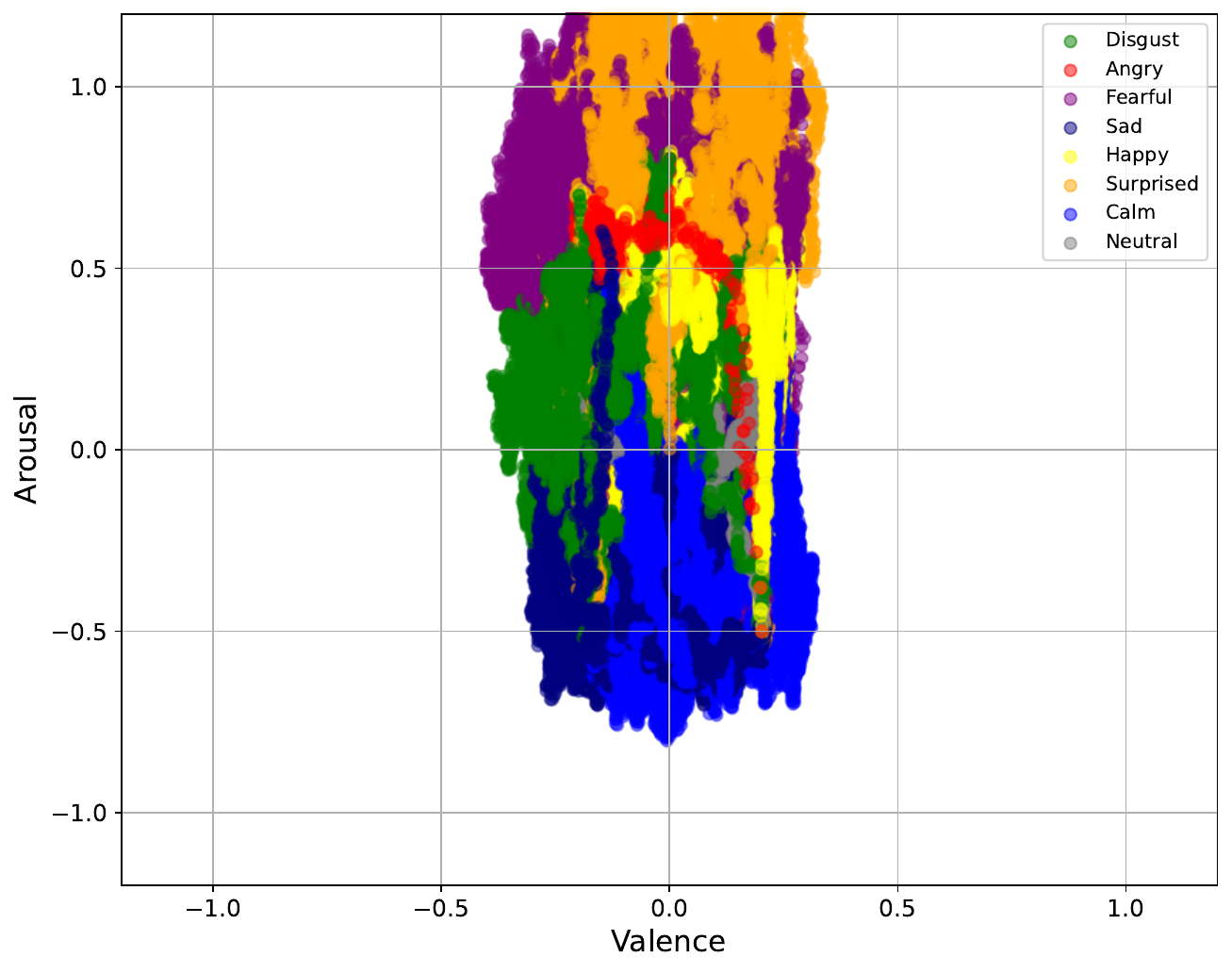}
    \caption{Low valence focus core affect}
    \label{fig:subfig3}
  \end{subfigure}
  \hfill
  \begin{subfigure}[b]{0.24\textwidth}
    \includegraphics[width=\linewidth]{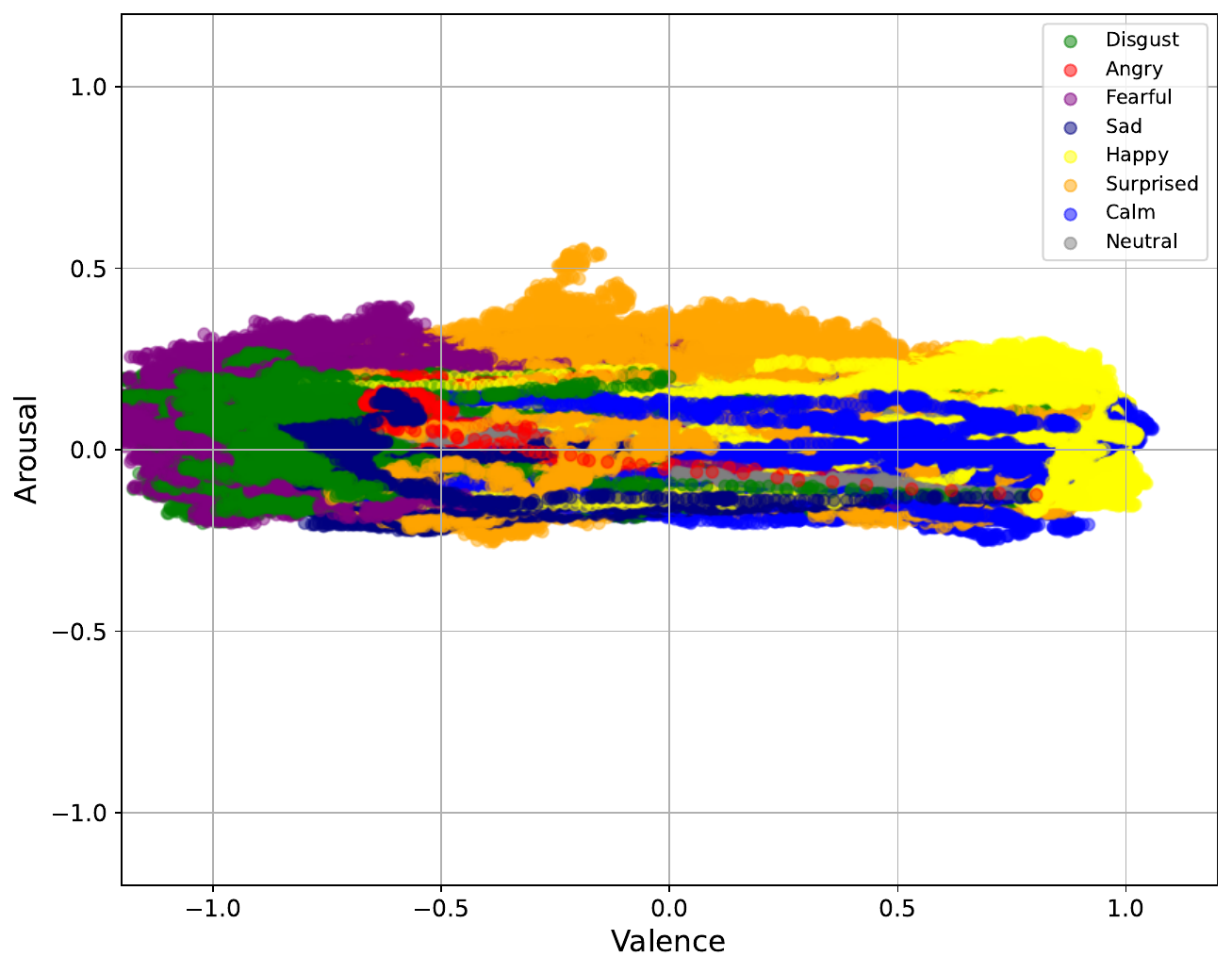}
    \caption{Low arousal focus core affect}
    \label{fig:subfig4}
  \end{subfigure}
  \caption{We used only Original core affect for Agent A, and each of these four core affect for Agent B in the experiment.Among them, Happy inverse core affect changes the mean of Happy from (0.9, 0.5) to (-0.9, -0.5), low valence focus core affect changes $\mu_v$, $\theta_v$, $\sigma_v$ to one-fourth of Original core affect, and low arousal focus core affect changes $\mu_a$, $\theta_a$, $\sigma_a$ to one-fourth of Original core affect.}
  \label{fig:four_figures}
\end{figure*}
We examine the effect of the presence or absence of MHNG, i.e., the interaction between two agents through the exchange of signs, on the formation of each agent's emotion category. Furthermore, by giving Agent B core affect that is different from that of Agent A, we consider the effect that differences in core affect have on the co-construction of emotions.

Specifically, we use the Adjusted Rand Index (ARI) \cite{hubert1985comparing} in two complementary ways. First, we measure ARI between the agent-emerged signs and the stimulus-level emotion category provided by RAVDESS (i.e., the emotion the actor was instructed to portray). We emphasise that this is not a supervised classification target—the model receives no label during learning—but rather a reference label that records which stimuli were intended to evoke the same emotion, against which we can assess whether the unsupervised co-construction process recovers psychologically meaningful clusters.
We use Cohen's Kappa coefficient \cite{cohen1960coefficient} as a complementary, chance-corrected measure of inter-agent agreement, and visualise the structure of the formed categories using heatmaps of recall against the RAVDESS reference labels.
Furthermore, we visualize the latent variable $z_d$ using PCA and t-distributed Stochastic Neighbor Embedding (t-SNE)~\cite{van2008visualizing}, and evaluate the structure of emotion categories by calculating the similarity of the latent space $z$ of the two agents using TopSim~\cite{kriegeskorte2008representational}. We also evaluate the differentiation of emotion categories using the Davies-Bouldin Score (DBS)~\cite{davies2009cluster}.

Based on our theoretical framework, we formally propose the following hypotheses:

\begin{enumerate}
    \item[\textbf{H1:}] \textbf{Emergent communication facilitates the co-construction of emotion.} We predict that the MHNG scenario will result in: (a) a stronger correspondence between learned categories and the RAVDESS reference categories, as measured by a higher ARI; (b) higher inter-agent agreement on emotional symbols (i.e., higher Kappa); and (c) more clearly differentiated individual emotion categories (i.e., lower DBS), compared to the No Communication scenario.
    
    \item[\textbf{H2:}] \textbf{Interoceptive similarity enhances the quality of co-constructed categories.} We predict that agents with similar core affect that communicate via MHNG will develop: (a) more accurate emotion categories (higher ARI); (b) more distinct emotion categories (lower DBS); and (c) more structurally similar latent representations (higher TopSim), compared to agents with divergent core affect models.
\end{enumerate}

\subsection{Dataset:Agent's body reactions as observations of the model}
When considering the emergence of emotion symbols, we use information based on the agent's own body reactions as the observation of the model. In other words, the agent's body reactions, such as facial expressions, vocal expressions, and interoception, when viewing a certain emotion evocative stimulus $E_d$ are treated as the input of the MVAE of each agent. 

We used the Ryerson Audio-Visual Database of Emotional Speech and Song (RAVDESS)~\cite{livingstone2018ryerson} as the facial and vocal expressions corresponding to this emotion evocative stimulus. RAVDESS is a video data set of speech and song in eight emotion categories (Neutral, Calm, Happy, Sad, Angry, Fearful, Disgust, Surprised) collected from 12 male and 12 female actors. 

In the experiment, we used only the speech video data, and obtained Facial Action Units (FAUs) at each time as facial expressions and Mel-frequency Cepstral Coefficients (MFCCs) as vocal expressions of agent. We used OpenFace~\cite{tadas2018openface}  for facial expressions. Using this method, the strengths of 35 FAUs were obtained for each frame of the video. In addition, 20-dimensional MFCC, $\Delta$MFCC, and $\Delta\Delta$MFCC were obtained from the video as vocal expressions. The frame length for obtaining MFCC was 2048 samples, and calculations were performed every 512 samples. 
\subsection{Generation of core affect}
\label{Generation of core affect}
\begin{figure}[htbp]
  \centering
  \includegraphics[width=0.48\textwidth]{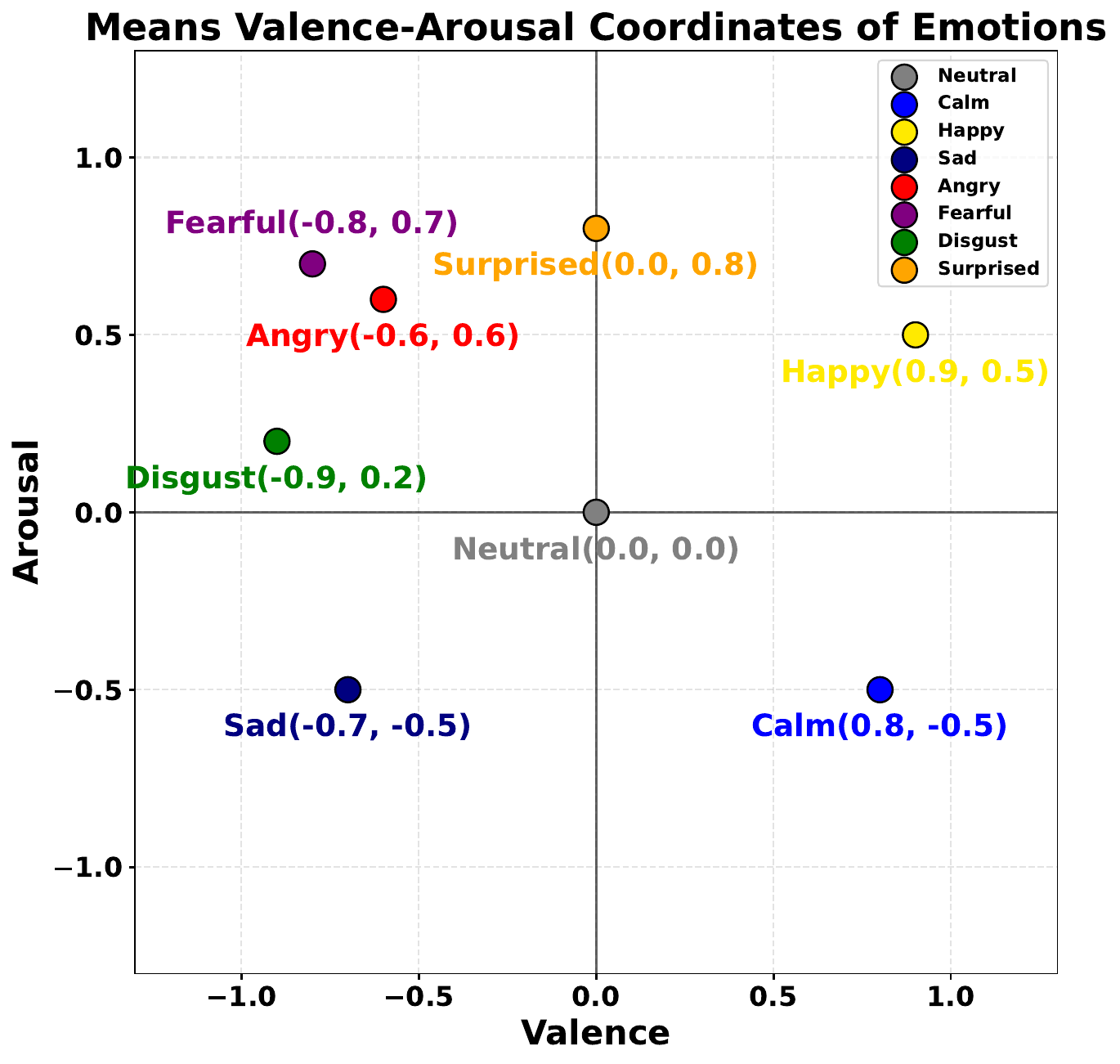}
  \caption{Mean valence–arousal coordinates for each emotion category. Colors are consistent with emotion labels}
  \label{fig:valence_arousal_means}
\end{figure}
\begin{figure}[htbp]
\centering
\begin{subfigure}[b]{0.45\textwidth}
    \centering
    \includegraphics[width=\textwidth]{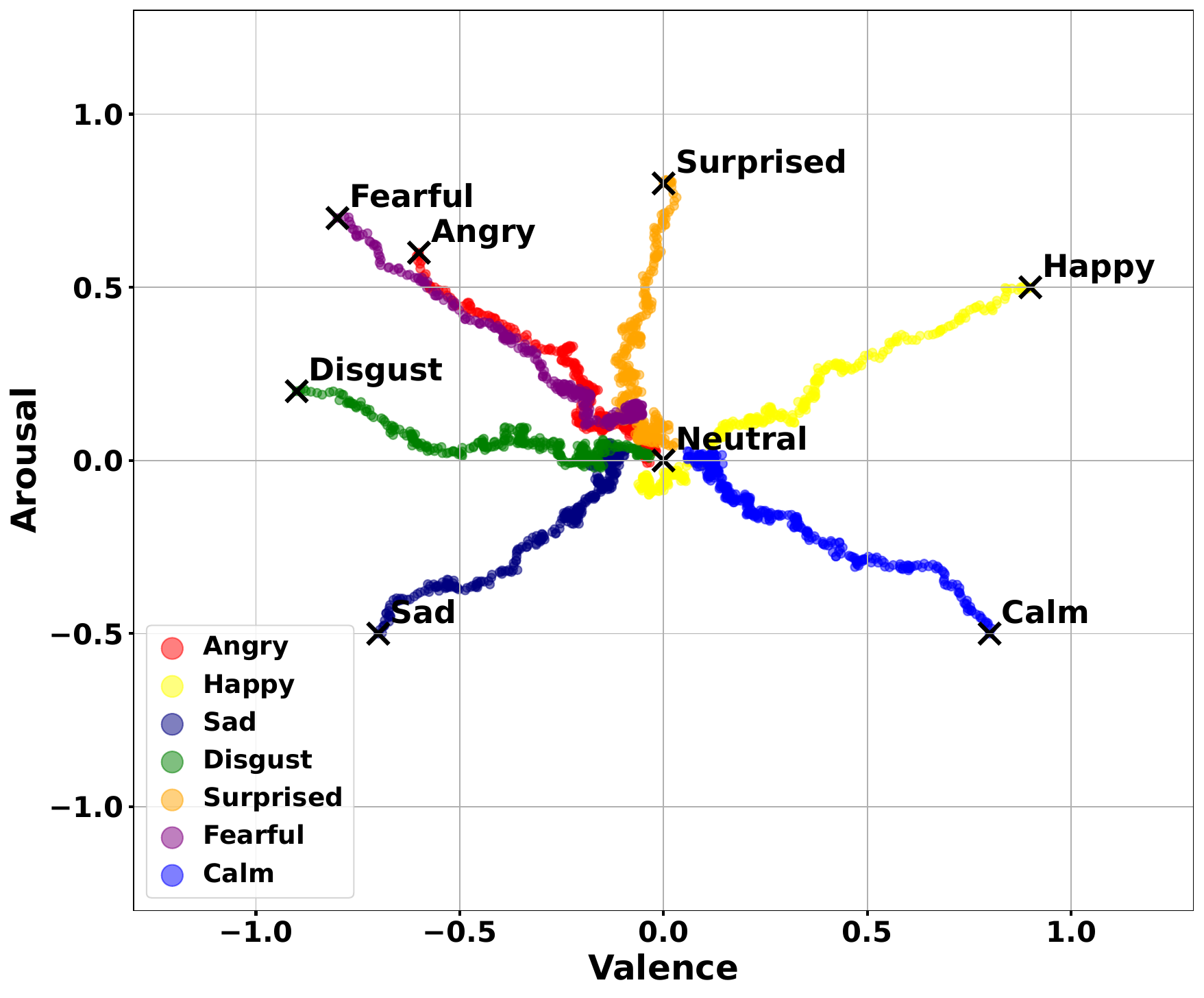}
    \caption{Target emotion: Neutral}
    \label{fig:target emotion is Neutral}
\end{subfigure}
\hfill
\begin{subfigure}[b]{0.45\textwidth}
    \centering
    \includegraphics[width=\textwidth]{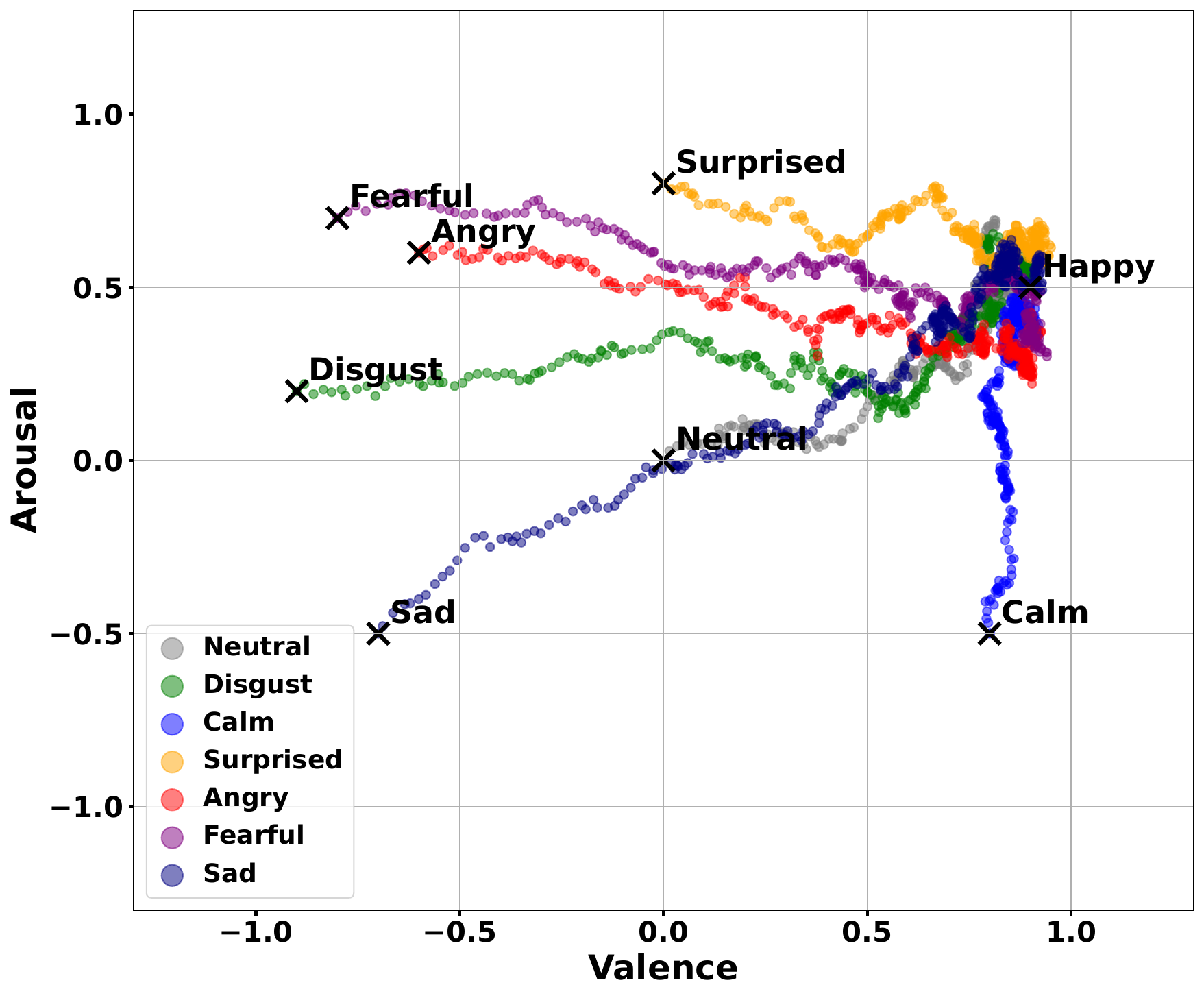}
    \caption{Target emotion: Happy}
    \label{fig:target emotion is Happy}
\end{subfigure}
\caption{Visualization of core affect trajectories generated by the Ornstein-Uhlenbeck stochastic process. The plots illustrate the stochastic nature of bodily reactions, showing temporal transitions from various initial states converging toward the mean vector of the target emotion ((a) Neutral and (b) Happy) in the Valence-Arousal space.}
\label{fig:overall_fig}
\end{figure}

We model core affect as a stochastic process to capture the inherent temporal fluctuations and physiological noise characteristic of bodily sensations. Specifically, we assume that core affect is governed by \textit{homeostatic regulation}—the tendency of the body's internal state to return to a baseline after being perturbed by stimuli. To mathematically represent this dynamics, we employ the Ornstein-Uhlenbeck (OU) process, which is a mean-reverting stochastic process defined as follows:
\begin{equation}
dX_t = \theta(\mu - X_t)\,dt + \sigma\,dW_t,
\end{equation}
where:

\begin{itemize}
  \item $X_t$: the value of the stochastic process at time $t$;
  \item $\theta > 0$: the mean reversion rate, which determines how quickly the process reverts to the mean $\mu$;
  \item $\mu$: the long-term mean toward which the process tends;
  \item $\sigma$: the volatility coefficient, controlling the magnitude of random fluctuations;
  \item $dt$: an infinitesimal time increment;
  \item $dW_t$: an infinitesimal increment of the Wiener process at time $t$.
\end{itemize}
In this process, the variable $X_t$ starts from a specific initial value and stochastically fluctuates, tending to revert towards the long-term mean $\mu$.
We referred to Russell’s Circumplex Model~\cite{russell1980circumplex}, which reduces core affect to two dimensions: valence and arousal, and set a target mean for each emotion. The target mean for each emotion is shown in Fig. \ref{fig:valence_arousal_means}.
We set the target mean of the OU process for each sample in the dataset to the mean vector corresponding to its labeled emotion. To simulate the ecological dynamics of emotional experience, we assume a sequential process where an agent transitions from a preceding emotional state to a new state triggered by a specific event. To capture these transition dynamics, each data sample is replicated seven times, with each replica initialized using the mean of one of the seven other emotion categories (different from the target). This initialization strategy effectively models the trajectory of core affect as it shifts from various prior states toward the new attractor defined by the current stimulus, thereby mitigating bias from any single fixed initial condition. Figs. \ref{fig:target emotion is Neutral} and \ref{fig:target emotion is Happy} illustrate examples where the target means correspond to Neutral and Happy, respectively.
The specific values for the mean reversion rate $\theta$ and volatility $\sigma$ assigned to each emotion category are detailed in Table~\ref{tab:interoception_set} in the Appendix.

\subsubsection{Asymmetric interoceptive profiles}
A central tenet of constructed emotion theory is that
interoceptive variability across individuals is the rule, not
the exception~\cite{barrett2017emotions}. Empirical work in
affective neuroscience supports this: interoceptive accuracy
modulates the intensity and quality of subjective emotional
experience~\cite{terasawa2013interoceptive}, and individuals
high in alexithymia show atypical interoceptive
sensitivity along the valence and arousal
dimensions~\cite{brewer2016alexithymia,murphy2017interoception}.
To probe whether MHNG-mediated communication can bridge such
heterogeneity, we construct three asymmetric core-affect
profiles in addition to the Original baseline (Fig.~3):
\begin{itemize}
\item \textbf{Happy-inverse}: the Happiness attractor is
inverted from $(0.9, 0.5)$ to $(-0.9, -0.5)$, simulating an
agent for whom positively valenced events elicit a negatively
valenced bodily response.
\item \textbf{Low-valence-focus}: $\mu_v$, $\theta_v$,
$\sigma_v$ are reduced to one-quarter of the Original values,
simulating an agent whose interoceptive readout along the
valence axis is attenuated.
\item \textbf{Low-arousal-focus}: $\mu_a$, $\theta_a$,
$\sigma_a$ are reduced to one-quarter, simulating an agent
with attenuated arousal sensitivity.
\end{itemize}
In all asymmetric experiments, Agent~A retains the Original
profile while Agent~B adopts one of the three asymmetric
profiles. We emphasise that these profiles are not intended
as models of pathological interoception; rather, they probe
the broader space of interoceptive variation that is normal
in any human population.

\subsection{Experimental setup}
To investigate the impact of communication within the framework of the co-construction of emotion, we investigate the latent variable $z_d$ of the GMM+MVAE for both agents in a model that adopts the MHNG framework for the Inter-GMM+MVAE (hereafter referred to as the MHNG scenario), a model in which the acceptance probability $r$ in MHNG is always 0 (hereafter referred to as the No communication scenario), and a model in which the acceptance probability $r$ is always 1 (hereafter referred to as the All-acceptance scenario). Here, the computational model for the No communication condition can be considered as each agent being equipped with its own GMM+MVAE.

    \begin{table*}[ht]
\centering
\caption{Clustering performance and quality metrics (mean ± std) under different emotional conditions and communication strategies. The top table displays the Adjusted Rand Index (ARI) and Cohen's Kappa, where higher values are better (↑). The bottom table shows the Davies-Bouldin Score (DBS), where lower values are better (↓), and Topic Similarity (TopSim), where higher values are better (↑). For all metrics, \textbf{bold} indicates the best result and \underline{underline} indicates the second-best.}
\label{tab:combined-results}
\setlength{\tabcolsep}{4pt}

\begin{tabular}{llccccccccccc}
\toprule
\multirow{3}{*}{Condition} & 
\multirow{3}{*}{} & 
\multicolumn{6}{c}{\textbf{ARI} (↑)} & 
\multicolumn{3}{c}{\textbf{Kappa} (↑)} \\
\cmidrule(lr){3-8} \cmidrule(lr){9-11}
& & \multicolumn{2}{c}{No Com.} & \multicolumn{2}{c}{MHNG} & \multicolumn{2}{c}{All Acc.} & No Com. & MHNG & All Acc. \\
& & Agent A & Agent B & Agent A & Agent B & Agent A & Agent B & & & \\
\midrule
Vision+Audio / Vision+Audio& 
& 0.14$\pm$0.03 & \underline{0.17$\pm$0.04} 
& \textbf{0.19$\pm$0.04} & \textbf{0.20$\pm$0.04} 
& \underline{0.17$\pm$0.07} & 0.15$\pm$0.07 
& -0.00$\pm$0.04 & \textbf{0.38$\pm$0.07} & \underline{0.34$\pm$0.11} \\
Original / Original & 
& \underline{0.28$\pm$0.08} & \underline{0.21$\pm$0.05} 
& \textbf{0.41$\pm$0.08} & \textbf{0.35$\pm$0.08} 
& 0.12$\pm$0.05 & 0.12$\pm$0.05 
& 0.01$\pm$0.06 & \textbf{0.51$\pm$0.09} & \underline{0.22$\pm$0.08} \\
Original / Happy inverse & 
& \underline{0.26$\pm$0.07} & \underline{0.23$\pm$0.05}
& \textbf{0.43$\pm$0.08} & \textbf{0.41$\pm$0.07} 
& 0.09$\pm$0.02 & 0.08$\pm$0.02 
& -0.01$\pm$0.03 & \textbf{0.49$\pm$0.07} & \underline{0.16$\pm$0.04} \\
Original / Low valence focus & 
& \textbf{0.30$\pm$0.08} & \underline{0.16$\pm$0.04} 
& \textbf{0.30$\pm$0.08} & \textbf{0.26$\pm$0.07} 
& \underline{0.06$\pm$0.01} & 0.07$\pm$0.02 
& -0.01$\pm$0.03 & \textbf{0.39$\pm$0.07} & \underline{0.14$\pm$0.02} \\
Original / Low arousal focus & 
& \underline{0.28$\pm$0.07} & \underline{0.09$\pm$0.01} 
& \textbf{0.34$\pm$0.07} & \textbf{0.20$\pm$0.04} 
& 0.02$\pm$0.01 & 0.03$\pm$0.01 
& -0.00$\pm$0.02 & \textbf{0.39$\pm$0.04} & \underline{0.06$\pm$0.02} \\
\bottomrule
\end{tabular}

\bigskip 

\begin{tabular}{llccccccccccc}
\toprule
\multirow{3}{*}{Condition} & 
\multirow{3}{*}{} & 
\multicolumn{6}{c}{\textbf{DBS} (↓)} & 
\multicolumn{3}{c}{\textbf{TopSim} (↑)} \\
\cmidrule(lr){3-8} \cmidrule(lr){9-11}
& & \multicolumn{2}{c}{No Com.} & \multicolumn{2}{c}{MHNG} & \multicolumn{2}{c}{All Acc.} & No Com. & MHNG & All Acc. \\
& & Agent A & Agent B & Agent A & Agent B & Agent A & Agent B & & & \\
\midrule
Vision+Audio / Vision+Audio& 
& \underline{4.03$\pm$0.68} & \underline{3.75$\pm$0.93} 
& \textbf{3.25$\pm$0.50} & \textbf{3.43$\pm$0.81} 
& 5.91$\pm$2.34 & 7.37$\pm$4.32 
& \underline{0.20$\pm$0.05} & 0.14$\pm$0.07 & \textbf{0.24$\pm$0.11} \\
Original / Original & 
& \underline{5.16$\pm$0.60} & \underline{5.44$\pm$1.11} 
& \textbf{4.41$\pm$0.85} & \textbf{4.47$\pm$0.80} 
& 10.86$\pm$2.90 & 11.07$\pm$2.73 
& 0.18$\pm$0.06 & \underline{0.22$\pm$0.06} & \textbf{0.25$\pm$0.08} \\
Original / Happy inverse & 
& \underline{4.28$\pm$0.63} & \underline{5.68$\pm$1.18} 
& \textbf{3.65$\pm$0.70} & \textbf{4.38$\pm$1.04} 
& 15.71$\pm$6.63 & 14.05$\pm$4.76 
& 0.12$\pm$0.06 & \textbf{0.16$\pm$0.07} & \underline{0.16$\pm$0.08} \\
Original / Low valence focus & 
& \underline{4.69$\pm$0.71} & \underline{6.34$\pm$1.32} 
& \textbf{4.39$\pm$0.84} & \textbf{5.44$\pm$1.01} 
& 13.86$\pm$3.57 & 16.51$\pm$6.33 
& 0.13$\pm$0.12 & \underline{0.15$\pm$0.06} & \textbf{0.25$\pm$0.04} \\
Original / Low arousal focus & 
& \underline{4.58$\pm$0.74} & \underline{7.34$\pm$1.69} 
& \textbf{4.03$\pm$0.71} & \textbf{6.03$\pm$1.08} 
& 32.99$\pm$10.27 & 29.66$\pm$6.67 
& 0.10$\pm$0.06 & \textbf{0.14$\pm$0.03} & \underline{0.12$\pm$0.05} \\
\bottomrule
\end{tabular}
\end{table*}
In this experiment, the data of 12 males from RAVDESS was used as the observation information for agent A, and the data of 12 females was used as the observation information for agent B. The number of input dimensions for each mode of MVAE is 35×109[frame] for $o_{vd}$, 3815 dimensions, 60×345[frame] for $o_{ad}$, 20700 dimensions, and 2×345[frames] for $o_{id}$, 690 dimensions. For RAVDESS video data with shorter video length, the data is padded with the last frame. The PoE is used for MVAE, and the number of dimensions of the latent variables in GMM+MVAE is set to 9. Experiments were performed 10 times for each scenario, and the results were evaluated using the ARI, DBS, heat map, Kappa coefficient, TopSim, and visualizations of the latent variables using PCA and t-SNE.

\begin{figure*}[t]
  \begin{center}
    \includegraphics[width=\linewidth]{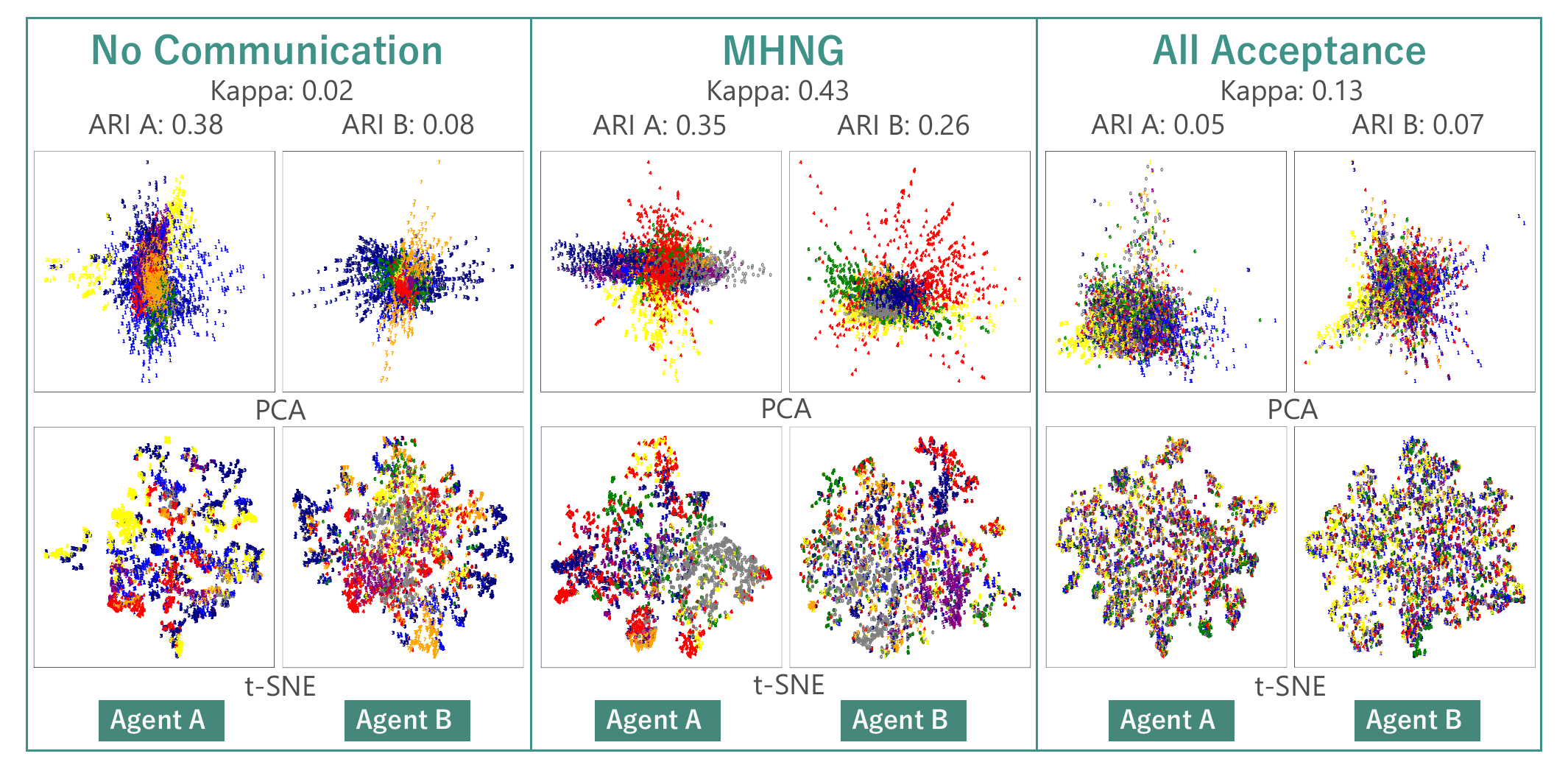}
    \caption{Result of PCA and t-SNE}
    \label{fig:Result of PCA and t-SNE}
  \end{center}
  \vspace{-1mm}
\end{figure*}

\begin{figure*}[htbp]
\centering
\begin{subfigure}[b]{0.45\textwidth}
    \centering
    \includegraphics[width=\textwidth]{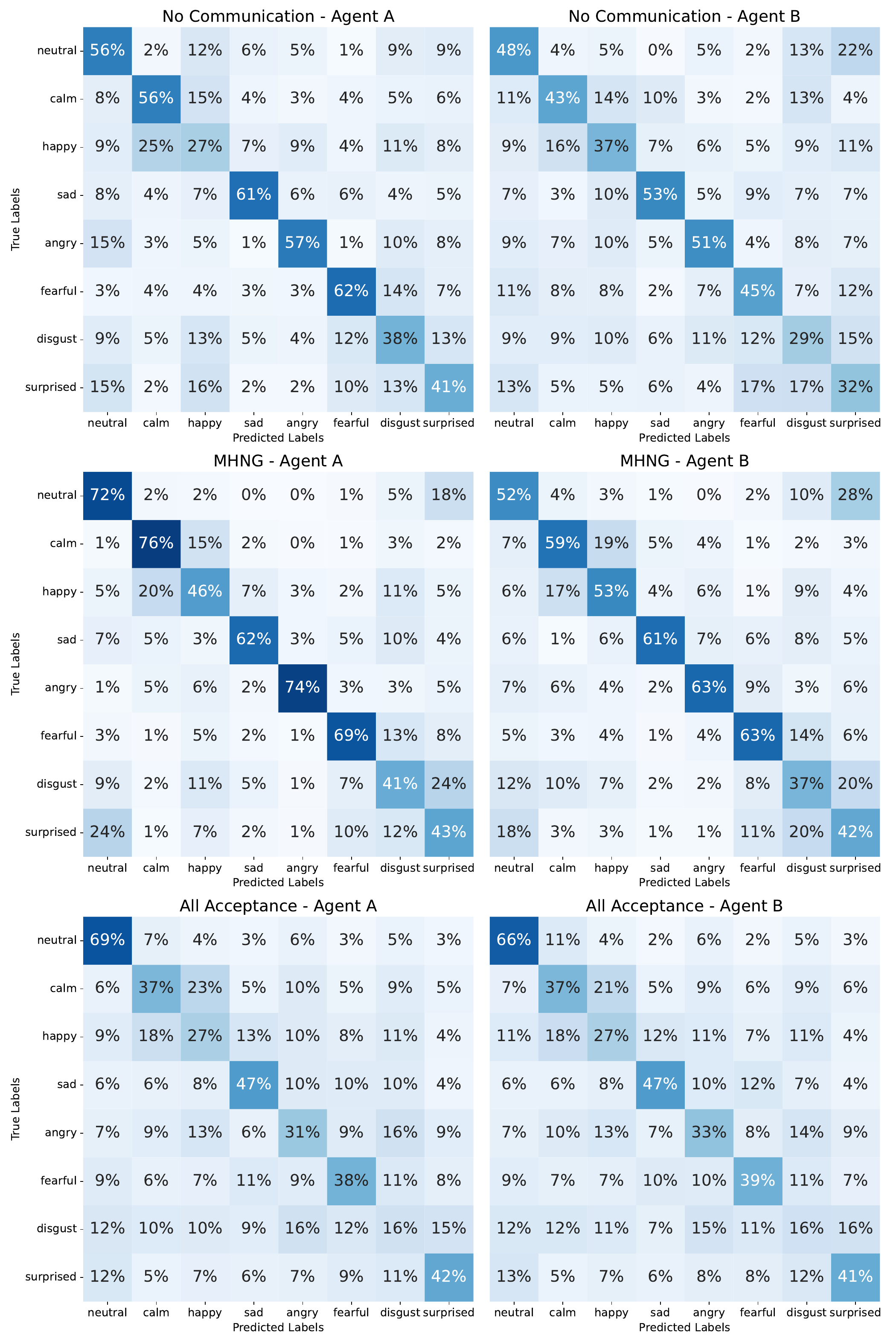} 
    \caption{Heat map of Original/Original}
    \label{fig:Original_heatmaps}
\end{subfigure}
\hfill
\begin{subfigure}[b]{0.45\textwidth}
    \centering
    \includegraphics[width=\textwidth]{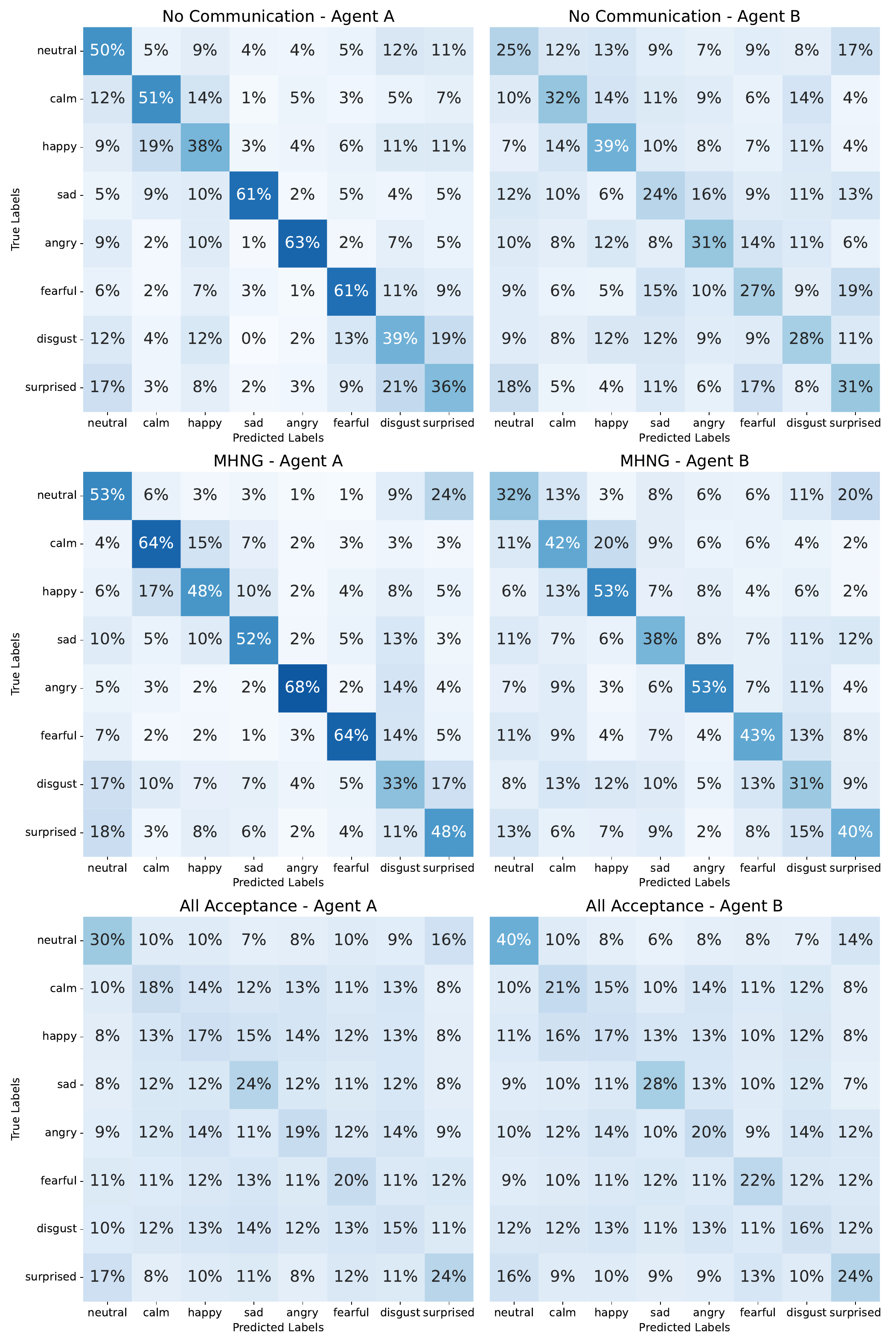} 
    \caption{Heat map of Original/Low arousal focus}
    \label{fig:Low arousal focus_heatmaps}
\end{subfigure}
\caption{The heat map uses recall to evaluate how well the model can recognize data with the same labels.}
\label{heatmaps}
\end{figure*}
\section{Experiment Results}

In this section, we showed the experimental results, organized around our primary research questions. We first analyze the general effect of emergent communication on the formation of emotion categories. We then examine the impact of communication on the underlying latent space structure. Finally, we investigate how divergence in core affect between agents affects the co-construction process.

\subsection{Effect of Emergent Communication on Category Formation}

To analysis the effect of emergent communication, we compared the three main experimental scenarios. The quantitative results, presented in Table~\ref{tab:combined-results}, show a clear performance hierarchy.

The \textbf{MHNG scenario} consistently outperformed the baselines across all key metrics. It achieved the highest ARI, indicating the strongest correspondence between the learned signs and the RAVDESS reference labels. Similarly, it yielded the highest Cohen's Kappa coefficient, signifying the strongest inter-agent agreement on the emerged emotional symbols. For the Davies-Bouldin Score (DBS), where lower values indicate better-defined and more separated clusters, the MHNG scenario produced the lowest (best) scores, as shown in Table~\ref{tab:combined-results}.

In contrast, the \textbf{No Communication} scenario performed moderately, serving as a baseline for independent learning. Notably, the \textbf{All-acceptance} scenario, which forces agents to accept all proposed symbols, consistently resulted in the poorest performance across ARI and DBS.

These findings suggest that emergent communication, when mediated by an intelligent and selective mechanism like MHNG, is highly effective for the co-construction of emotion categories. It not only helps agents to differentiate their own emotional concepts more clearly (lower DBS) but also enables them to converge on a shared, accurate symbolic system (higher ARI and Kappa). The failure of the All-acceptance scenario highlights that mere information exchange is insufficient; the ability to selectively reject incongruent symbols is crucial for robust social learning.

\subsection{Analysis of Latent Space Structure}

We next investigated whether communication fundamentally reshaped the agents' overall internal representations or primarily aligned their symbolic labels. Quantitative analysis of the latent space using TopSim (Table~\ref{tab:combined-results}) revealed no significant difference in structural similarity between the MHNG and No Communication scenarios. The PCA plots in Figure~\ref{fig:Result of PCA and t-SNE} visually corroborate this finding, revealing no significant differences in the global data structure across the three scenarios.

However, a more fine-grained visualization using t-SNE (Figure~\ref{fig:Result of PCA and t-SNE}) suggests a subtle but important effect. In the MHNG scenario, some clusters (e.g., the purple and gray clusters) appear more compact and distinctly formed between the two agents compared to the No Communication scenario.

Taken together, the primary influence of MHNG is not on fundamentally reshaping the agents' latent spaces ($z_d$), which remain heavily grounded in their own sensory experience. Because the agents exchange only a single, low-bandwidth discrete sign ($w_d$) for each emotional stimulus—not the rich, high-dimensional sequence data—it is logical that the main impact would be on the symbolic mapping rather than a fundamental reorganization of the underlying latent space. Consequently, this process establishes a shared symbolic layer upon individual, subjective experiences, facilitating mutual understanding without requiring identical internal states.

\subsection{Effect of Interoceptive Divergence}
We next explored the role of interoceptive similarity in the co-construction process by introducing systematic divergence in the agents' core affect. As a baseline, the introduction of any core affect yielded substantially better ARI and Kappa scores than experiments relying on vision and audio alone (Table~II, Vision+Audio row vs.\ all Original-paired rows), underscoring the importance of interoception for emotion categorization. Among the four MHNG configurations involving core affect, the symmetric Original/Original condition yielded the highest inter-agent agreement (Kappa $=0.51$).

The most insightful results, however, emerged from the asymmetric conditions, in which the two agents have systematically different interoceptive dynamics. We focus on the ``Original / Low arousal focus'' condition, where Agent~A uses the Original core affect and Agent~B uses a core affect whose arousal-axis parameters $(\mu_a,\theta_a,\sigma_a)$ are attenuated to one quarter of the Original values. The recall heatmaps in Fig.\ref{heatmaps} reveal two distinct patterns of categorical reshaping under MHNG.
\begin{enumerate}
    \item \textbf{Observation 1: Categorical structure of the agent with attenuated arousal sensitivity is sharpened under MHNG:} Comparing the No~Communication and MHNG conditions for Agent~B in Fig.~7b, the diagonal recall values increase for all eight emotion categories: Neutral $25\% \to 32\%$, Calmness $32\% \to 42\%$, Happiness $39\% \to 53\%$, Sadness $24\% \to 38\%$, Anger $31\% \to 53\%$, Fear $27\% \to 43\%$, Disgust $28\% \to 31\%$, and Surprise $31\% \to 40\%$. The largest gains occur for Anger ($+22$ points) and Fear ($+16$ points), the two categories whose RAVDESS reference labels lie at the high-arousal end of the valence-arousal space (cf.\ Fig.~4) and whose interoceptive signatures are therefore most affected by Agent~B's reduced arousal sensitivity. The overall sharpening is also reflected in the corresponding ARI for Agent~B in the Original/Low arousal focus row of Table~II, which improves from $0.09 \pm 0.01$ (No~Com.) to $0.20 \pm 0.04$ (MHNG). MHNG-mediated communication thus enables Agent~B---whose interoceptive readout provides only weakly differentiated arousal information---to develop categorical assignments that more closely correspond to the RAVDESS reference labels.
    \item \textbf{Observation 2: Categorical boundaries of the other agent are reshaped, not uniformly improved:} The change in Agent~A's recall pattern under MHNG is not a uniform improvement or degradation but a category-specific reorganisation. Comparing Agent~A in Fig.~7a (symmetric Original/Original under MHNG) and Fig.~7b (asymmetric Original/Low arousal focus under MHNG), the diagonal recall values move in opposite directions for different categories: six categories show reduced diagonal recall (Neutral $72\% \to 53\%$, Calmness $76\% \to 64\%$, Sadness $62\% \to 52\%$, Anger $74\% \to 68\%$, Fear $69\% \to 64\%$, Disgust $41\% \to 33\%$), while two show small increases (Happiness $46\% \to 48\%$, Surprise $43\% \to 48\%$). The corresponding ARI for Agent~A in Table~II shifts modestly, from $0.41 \pm 0.08$ in Original/Original to $0.34 \pm 0.07$ in Original/Low arousal focus.
\end{enumerate}

Together, Observations~1 and~2 show that asymmetric interoception leads not to a one-directional transfer of structure from one agent to the other, but to a \emph{category-specific bidirectional reshaping} of both agents' categorical systems. Even under such reshaping, the inter-agent Kappa under MHNG remains high ($0.39 \pm 0.04$ for Original/Low arousal focus, comparable to the other asymmetric MHNG conditions in Table~II), indicating that the two agents still converge on a shared symbolic system. The interpretation of these reshaping patterns---specifically, the constructionist reading that interoceptive heterogeneity is a constitutive feature of emotional life rather than a deficit to be corrected---is developed in Section~\ref{sec:discussion-heterogeneity}.

\section{Discussion}
The experimental results in Section~V can be situated in three
broader theoretical contexts: the dissociation between symbolic
and perceptual layers in co-construction, the role of interoceptive
heterogeneity in emotion formation, and the implications for
cognitive developmental robotics and symbol emergence in robotics.

\subsection{MHNG operates at the symbolic, not the perceptual layer}

Across all conditions, MHNG-mediated communication left the
structural similarity (TopSim) of the two agents' latent spaces
$z^{A}_{d}$ and $z^{B}_{d}$ essentially unchanged
(Section~V-2), while substantially improving inter-agent agreement
at the symbolic level $w_{d}$ (Kappa rose from $0.01$ to $0.51$
in the Original/Original condition; Table~II). This dissociation
is by design: for each stimulus the speaker transmits only a
single integer index $w_{d} \in \{1,\ldots,K\}$ ($K=9$ in our
experiments, i.e.\ $\log_{2}9 \approx 3.2$ bits per stimulus),
not the full multimodal observation tensor (vision: 3{,}815-dim,
audio: 20{,}700-dim, interoception: 690-dim, totalling
$\approx 25{,}205$ dimensions per sample). Because the channel
transmits only this minimal categorical information, MHNG can
only modify the agents' categorical assignments and the GMM
cluster parameters, not the continuous, modality-grounded latent
geometry produced by the MVAE encoders. Consequently,
co-construction establishes a shared symbolic layer $w$ on top
of individually embodied, subjective continuous representations
$z$. This is precisely the architectural property that enables
mutual understanding without requiring agents to share identical
internal states or perceptual structures---a computational
analogue of the long-standing observation in linguistics that
words are public while meanings are private \cite{quine1960word}.

\subsection{Co-construction does not require interoceptive isomorphism}\label{sec:discussion-isomorphism}

Across the three asymmetric conditions
(Original/Happy-inverse, Original/Low-valence-focus,
Original/Low-arousal-focus), MHNG produced inter-agent Kappa
values of $0.39$--$0.49$, comparable to the symmetric
Original/Original baseline (Kappa $=0.51$) and dramatically
higher than the No Communication ($\approx 0$) and All
Acceptance ($\le 0.22$) baselines. This computationally
instantiates Gendron and Barrett's theoretical claim that
emotion perception relies on \emph{conceptual synchrony}
rather than physiological mirroring~\cite{gendron2018emotion}:
even when two agents process the same stimulus through
systematically different interoceptive dynamics, communication
enables them to converge on a shared categorical structure.

Crucially, this convergence is not enforced uniformity. The
All Acceptance condition---in which one agent unconditionally
adopts the other's signs---yields \emph{worse} alignment with
the reference labels (ARI $\le 0.17$ in all conditions) and
\emph{worse} cluster quality (DBS $\ge 10$) than No
Communication. The selectivity of the Metropolis--Hastings
rejection step is what allows each agent to retain category
boundaries that respect its own interoceptive evidence while
still aligning categorically with its partner. In the
constructed-emotion framework, this maps onto the distinction
between \emph{shared categorical knowledge} and \emph{private
embodied experience}: two interlocutors can agree on the
applicability of the word ``anger'' without having identical
visceral reactions to it.

\subsection{Interoceptive heterogeneity as a feature, not a defect}\label{sec:discussion-heterogeneity}

Real human interoception is heterogeneous: individuals differ
in interoceptive accuracy, attention, and valence/arousal
sensitivity, with such differences linked to traits ranging
from anxiety to alexithymia and contemplative
expertise~\cite{terasawa2013interoceptive,brewer2016alexithymia,
murphy2017interoception}. The constructed-emotion framework
treats this heterogeneity as constitutive of, rather than
noise around, emotional life~\cite{barrett2017emotions}.

Our results align with this constitutive view in three ways.
\emph{First}, an agent learning alone (No Com.) only loosely
recovers the experimenter-defined emotion structure
(ARI $\approx 0.21$--$0.30$; Table~II), suggesting that no
single body-grounded model is sufficient. \emph{Second}, when
two agents with non-identical interoceptive profiles
communicate via MHNG, their joint categorical system reaches
a higher Kappa and ARI than either does alone---suggesting
that interoceptive diversity is itself a resource for richer
category formation, not an obstacle to overcome.
\emph{Third}, the asymmetric reshaping of recall patterns
under asymmetric conditions (e.g., the
Original/Low-arousal-focus heatmap in Fig.~7b) is reminiscent
of how, in everyday social interaction, interlocutors shift
their emotional vocabulary depending on whom they are
speaking with---a flexibility that is hard to explain on
essentialist (e.g., basic-emotion) accounts but natural under
constructionist views.

It should be emphasised that the asymmetric conditions in our
experiments are not intended as models of pathological
interoception; rather, they probe the broader space of
interoceptive variation that is normal in any human
population. The corresponding categorical reshaping observed
under MHNG should therefore be read as a model of \emph{how
diverse bodies arrive at shared concepts}, not as a model of
``correcting'' a deficient agent.

\subsection{Implications for cognitive developmental robotics
and symbol emergence in robotics}

From a cognitive developmental robotics
perspective~\cite{asada2015towards,asada2016modeling},
our results suggest a concrete mechanism by which an
artificial agent can acquire human-aligned emotion concepts
without requiring its body to faithfully replicate human
physiology. Prior CDR work on affective development modelled
emotion formation within a single agent~\cite{Horii:2018,
hieida2018deepemotioncomputationalmodel}; the present work extends this to the
\emph{social loop}, showing that two agents with different
embodiments can still converge on a shared emotional
vocabulary through naming-game-like interaction. Such a
mechanism is a candidate building block for caregiver--infant
emotional learning models, where caregiver and infant clearly
do not share identical interoceptive states.

From a symbol emergence in robotics
perspective~\cite{taniguchi2016symbol,taniguchi2024cpc},
the present work extends the
Inter-GMM+MVAE framework---originally validated on physical
objects~\cite{hoang10emergent}---to the more abstract,
body-grounded domain of emotion. The success of this extension
is non-trivial: emotion categories lack the stable
visual/physical regularities that ground object names, and yet
MHNG still recovers shared categorical structure. This
suggests that the symbol emergence framework is not restricted
to perceptually grounded categories but can extend to
internally grounded ones, opening a path toward modelling the
emergence of social, evaluative, and abstract concepts
in artificial agents.

\subsection{Limitations and future work}

Our model represents a deliberate simplification of real
emotional co-construction, and four limitations are worth
flagging.

\paragraph{Stimulus ecology}
RAVDESS contains posed performances by professional actors,
which are known to differ from spontaneous expressions in
temporal dynamics and feature
distribution~\cite{calvo2016perceptual}; ecological validity
is thus limited.

\paragraph{Synthetic interoception}
Core affect is simulated via an Ornstein--Uhlenbeck process
rather than measured from physiological signals such as
heart-rate variability or galvanic skin response. Integrating
real interoceptive measurements is a natural next step.

\paragraph{Channel bandwidth}
The communication channel transmits a single discrete sign
per stimulus, whereas human emotional communication is
continuous and multimodal---facial expressions, prosody, and
gesture all carry affective signal. Extending the present
framework to multimodal sign exchange (e.g., pairs of agents
exchanging facial-expression-like vectors as well as discrete
category labels) is an important direction for future work,
as it would bring the model closer to the rich semiotic
exchange characteristic of human emotional interaction.

\paragraph{Population scale}
The system uses two agents only; cultural-level emotion norms
emerge in populations of many interacting agents, an extension
that is straightforward in principle within the CPC framework.
Future work will scale the system to larger groups to
investigate how shared emotional vocabularies stabilise into
culture-level norms.

\section{Conclusion}

This study presented a computational instantiation of the
co-construction of emotion, using the Inter-GMM+MVAE framework
grounded in Collective Predictive Coding to simulate how two
embodied agents form and align emotion categories from
multimodal sensory experience.

Three findings stand out. (i)~Selective communication via MHNG significantly improves inter-agent agreement (Kappa) and clarity (DBS) of the emerged emotion categories, while non-selective communication degrades performance. (ii)~Communication operates primarily at the symbolic layer ($w_{d}$), leaving the modality-grounded latent geometry ($z_{d}$) largely intact ---the key property that enables agents with different embodiments to share categories without sharing internal states.
(iii)~Asymmetric interoceptive profiles do not prevent co-construction; instead, they yield distinct, category-specific reshaping patterns that are consistent with the constructed-emotion view of interoceptive heterogeneity as constitutive of emotional life.

To our knowledge, this work provides the first computational validation of the co-constructionist view of emotion perception, and extends the applicability of the CPC framework from physical objects to the abstract, socially-grounded domain of human emotion. Future work will (a)~replace simulated core affect with empirically measured physiological signals, (b)~extend the communication channel from a single discrete sign to multimodal signals such as facial expressions and prosody, and (c)~scale beyond two agents to populations of interacting agents, enabling investigation of how culture-level emotional norms might emerge from local CPC dynamics.

\appendices
\section{VAE Architecture}


\begin{table}[htbp]
\centering
\caption{The parameters of each emotion used to generate core affect}
\label{tab:interoception_set}
\begin{tabular}{lcccccc}
\hline
\textbf{Emotion} & $\mu_V$ & $\mu_A$ & $\sigma_V$ & $\sigma_A$ & $\theta_V$ & $\theta_A$ \\
\hline
Neutral    &  0.00 &  0.00  & 0.090  & 0.090  & 1.5  & 1.5  \\
Calm       &  0.80 & -0.50  & 0.135  & 0.180  & 2.1  & 1.8  \\
Happy      &  0.90 &  0.50  & 0.090  & 0.225  & 2.7  & 2.4  \\
Sad        & -0.70 & -0.50  & 0.180  & 0.135  & 2.4  & 2.1  \\
Angry      & -0.60 &  0.60  & 0.225  & 0.270  & 1.8 & 2.7 \\
Fearful    & -0.80 &  0.70  & 0.270  & 0.315  & 1.5  & 3.0  \\
Disgust    & -0.90 &  0.20  & 0.225  & 0.225  & 2.1  & 2.4  \\
Surprised  &  0.00 &  0.80  & 0.180  & 0.360  & 1.2  & 1.8  \\
\hline
\end{tabular}

\end{table}



\ifCLASSOPTIONcaptionsoff
  \newpage
\fi

\bibliographystyle{IEEEtran}
\bibliography{IEEEabrv,Bibliography}

\begin{thebibliography}{10}
\providecommand{\url}[1]{#1}
\csname url@rmstyle\endcsname
\providecommand{\newblock}{\relax}
\providecommand{\bibinfo}[2]{#2}
\providecommand\BIBentrySTDinterwordspacing{\spaceskip=0pt\relax}
\providecommand\BIBentryALTinterwordstretchfactor{4}
\providecommand\BIBentryALTinterwordspacing{\spaceskip=\fontdimen2\font plus
\BIBentryALTinterwordstretchfactor\fontdimen3\font minus \fontdimen4\font\relax}
\providecommand\BIBforeignlanguage[2]{{%
\expandafter\ifx\csname l@#1\endcsname\relax
\typeout{** WARNING: IEEEtran.bst: No hyphenation pattern has been}%
\typeout{** loaded for the language `#1'. Using the pattern for}%
\typeout{** the default language instead.}%
\else
\language=\csname l@#1\endcsname
\fi
#2}}

\bibitem{Lazarus1991EmotionAndAdaptation}
R.~S. Lazarus, \emph{Emotion and Adaptation}.\hskip 1em plus 0.5em minus 0.4em\relax Oxford University Press, 1991.

\bibitem{Scherer2005WhatAreEmotions}
K.~R. Scherer, ``What are emotions? and how can they be measured?'' \emph{Social Science Information}, vol.~44, no.~4, pp. 695--729, 2005.

\bibitem{Damasio1999TheFeelingOfWhatHappens}
A.~R. Damasio, \emph{The Feeling of What Happens: Body and Emotion in the Making of Consciousness}.\hskip 1em plus 0.5em minus 0.4em\relax Harcourt Brace, 1999.

\bibitem{russell1980circumplex}
J.~A. Russell, ``A circumplex model of affect.'' \emph{Journal of personality and social psychology}, vol.~39, no.~6, p. 1161, 1980.

\bibitem{Mesquita1992CulturalVariationsInEmotions}
B.~Mesquita and N.~H. Frijda, ``Cultural variations in emotions: A review,'' \emph{Psychological Bulletin}, vol. 112, no.~2, pp. 179--204, 1992.

\bibitem{Kitayama1994EmotionAndCulture}
S.~Kitayama and H.~R. Markus, \emph{Emotion and Culture: Empirical Studies of Mutual Influence}.\hskip 1em plus 0.5em minus 0.4em\relax American Psychological Association, 1994.

\bibitem{barrett2017emotions}
L.~F. Barrett, \emph{How emotions are made: The secret life of the brain}.\hskip 1em plus 0.5em minus 0.4em\relax Pan Macmillan, 2017.

\bibitem{gendron2018emotion}
M.~Gendron and L.~F. Barrett, ``Emotion perception as conceptual synchrony,'' \emph{Emotion Review}, vol.~10, no.~2, pp. 101--110, 2018.

\bibitem{Damasio1994DescartesError}
A.~R. Damasio, \emph{Descartes' Error: Emotion, Reason, and the Human Brain}.\hskip 1em plus 0.5em minus 0.4em\relax G. P. Putnam's Sons, 1994.

\bibitem{Ekman1992AnArgumentForBasicEmotions}
P.~Ekman, ``An argument for basic emotions,'' \emph{Cognition and Emotion}, vol.~6, no. 3-4, pp. 169--200, 1992.

\bibitem{lindquist2012brain}
K.~A. Lindquist, T.~D. Wager, H.~Kober, E.~Bliss-Moreau, and L.~F. Barrett, ``The brain basis of emotion: a meta-analytic review,'' \emph{Behavioral and brain sciences}, vol.~35, no.~3, pp. 121--143, 2012.

\bibitem{barrett2006emotions}
L.~F. Barrett, ``Are emotions natural kinds?'' \emph{Perspectives on psychological science}, vol.~1, no.~1, pp. 28--58, 2006.

\bibitem{russell1994there}
J.~A. Russell, ``Is there universal recognition of emotion from facial expression? a review of the cross-cultural studies.'' \emph{Psychological bulletin}, vol. 115, no.~1, p. 102, 1994.

\bibitem{gendron2014perceptions}
M.~Gendron, D.~Roberson, J.~M. van~der Vyver, and L.~F. Barrett, ``Perceptions of emotion from facial expressions are not culturally universal: evidence from a remote culture.'' \emph{Emotion}, vol.~14, no.~2, p. 251, 2014.

\bibitem{Friston:2017}
K.~Friston, T.~FitzGerald, F.~Rigoli, P.~Schwartenbeck, and G.~Pezzulo, ``Active inference: a process theory,'' \emph{Neural Computation}, vol.~29, no.~1, pp. 1--49, 2017.

\bibitem{seth2016active}
A.~K. Seth and K.~J. Friston, ``Active interoceptive inference and the emotional brain,'' \emph{Philosophical Transactions of the Royal Society B: Biological Sciences}, vol. 371, no. 1708, p. 20160007, 2016.

\bibitem{Horii:2018}
T.~Horii, Y.~Nagai, and M.~Asada, ``Modeling development of multimodal emotion perception guided by tactile dominance and perceptual improvement,'' \emph{IEEE Transactions on Cognitive and Developmental Systems}, vol.~10, no.~3, pp. 762--775, 2018.

\bibitem{hieida2018deepemotioncomputationalmodel}
\BIBentryALTinterwordspacing
C.~Hieida, T.~Horii, and T.~Nagai, ``Deep emotion: A computational model of emotion using deep neural networks,'' 2018. [Online]. Available: \url{https://arxiv.org/abs/1808.08447}
\BIBentrySTDinterwordspacing

\bibitem{taniguchi2016symbol}
T.~Taniguchi \emph{et~al.}, ``Symbol emergence in robotics: a survey,'' \emph{Advanced Robotics}, vol.~30, no. 11-12, pp. 706--728, 2016.

\bibitem{hagiwara2019symbol}
Y.~Hagiwara, H.~Kobayashi, A.~Taniguchi, and T.~Taniguchi, ``Symbol emergence as an interpersonal multimodal categorization,'' \emph{Frontiers in Robotics and AI}, vol.~6, p. 134, 2019.

\bibitem{taniguchi2024cpc}
T.~Taniguchi, ``Collective predictive coding hypothesis: symbol emergence as decentralized bayesian inference,'' \emph{Frontiers in Robotics and AI}, vol.~11, p. 1353870, 2024.

\bibitem{hoang10emergent}
N.~L. Hoang, T.~Taniguchi, Y.~Hagiwara, and A.~Taniguchi, ``Emergent communication of multimodal deep generative models based on metropolis-hastings naming game,'' \emph{Frontiers in Robotics and AI}, vol.~10, 2024.

\bibitem{seth2013interoceptive}
A.~K. Seth, ``Interoceptive inference, emotion, and the embodied self,'' \emph{Trends in cognitive sciences}, vol.~17, no.~11, pp. 565--573, 2013.

\bibitem{taniguchi2023emergent}
T.~Taniguchi, Y.~Yoshida, Y.~Matsui, N.~Le~Hoang, A.~Taniguchi, and Y.~Hagiwara, ``Emergent communication through metropolis-hastings naming game with deep generative models,'' \emph{Advanced Robotics}, vol.~37, no.~19, pp. 1266--1282, 2023.

\bibitem{hagiwara2022multiagent}
Y.~Hagiwara, K.~Furukawa, A.~Taniguchi, and T.~Taniguchi, ``Multiagent multimodal categorization for symbol emergence: emergent communication via interpersonal cross-modal inference,'' \emph{Advanced Robotics}, vol.~36, no. 5-6, pp. 239--260, 2022.

\bibitem{sakurai2026mh}
K.~Sakurai, H.~Uenoyama, A.~Taniguchi, and T.~Taniguchi, ``Mh-mug: Collaborative music generation game between ai agents towards emergent musical creativity,'' \emph{IEEE Access}, 2026.

\bibitem{wu2018multimodal}
M.~Wu and N.~Goodman, ``Multimodal generative models for scalable weakly-supervised learning,'' \emph{Advances in neural information processing systems}, vol.~31, 2018.

\bibitem{shi2019variational}
Y.~Shi, B.~Paige, P.~Torr, \emph{et~al.}, ``Variational mixture-of-experts autoencoders for multi-modal deep generative models,'' \emph{Advances in neural information processing systems}, vol.~32, 2019.

\bibitem{sutter2021generalized}
T.~M. Sutter, I.~Daunhawer, and J.~E. Vogt, ``Generalized multimodal elbo,'' \emph{arXiv preprint arXiv:2105.02470}, 2021.

\bibitem{hubert1985comparing}
L.~Hubert and P.~Arabie, ``Comparing partitions,'' \emph{Journal of classification}, vol.~2, pp. 193--218, 1985.

\bibitem{cohen1960coefficient}
J.~Cohen, ``A coefficient of agreement for nominal scales,'' \emph{Educational and psychological measurement}, vol.~20, no.~1, pp. 37--46, 1960.

\bibitem{van2008visualizing}
L.~Van~der Maaten and G.~Hinton, ``Visualizing data using t-sne.'' \emph{Journal of machine learning research}, vol.~9, no.~11, 2008.

\bibitem{kriegeskorte2008representational}
N.~Kriegeskorte, M.~Mur, and P.~A. Bandettini, ``Representational similarity analysis-connecting the branches of systems neuroscience,'' \emph{Frontiers in systems neuroscience}, vol.~2, p. 249, 2008.

\bibitem{davies2009cluster}
D.~L. Davies and D.~W. Bouldin, ``A cluster separation measure,'' \emph{IEEE transactions on pattern analysis and machine intelligence}, no.~2, pp. 224--227, 2009.

\bibitem{livingstone2018ryerson}
S.~R. Livingstone and F.~A. Russo, ``The ryerson audio-visual database of emotional speech and song (ravdess): A dynamic, multimodal set of facial and vocal expressions in north american english,'' \emph{PloS one}, vol.~13, no.~5, p. e0196391, 2018.

\bibitem{tadas2018openface}
B.~Tadas, Z.~Amir, L.~Y. Chong, and M.~Louis-Philippe, ``Openface 2.0: Facial behavior analysis toolkit,'' in \emph{13th IEEE International Conference on Automatic Face \& Gesture Recognition}, 2018.

\bibitem{terasawa2013interoceptive}
Y.~Terasawa, H.~Fukushima, and S.~Umeda, ``How does interoceptive awareness interact with the subjective experience of emotion? an fmri study,'' \emph{Human Brain Mapping}, vol.~34, no.~3, pp. 598--612, 2013.

\bibitem{brewer2016alexithymia}
R.~Brewer, R.~Cook, and G.~Bird, ``Alexithymia: a general deficit of interoception,'' \emph{Royal Society Open Science}, vol.~3, no.~10, p. 150664, 2016.

\bibitem{murphy2017interoception}
J.~Murphy, R.~Brewer, C.~Catmur, and G.~Bird, ``Interoception and psychopathology: A developmental neuroscience perspective,'' \emph{Developmental Cognitive Neuroscience}, vol.~23, pp. 45--56, 2017.

\bibitem{quine1960word}
W.~V.~O. Quine, \emph{Word and Object}.\hskip 1em plus 0.5em minus 0.4em\relax MIT Press, 1960.

\bibitem{asada2015towards}
M.~Asada, ``Towards artificial empathy,'' \emph{International Journal of Social Robotics}, vol.~7, no.~1, pp. 19--33, 2015.

\bibitem{asada2016modeling}
------, ``Modeling early vocal development through infant--caregiver interaction,'' \emph{IEEE Transactions on Cognitive and Developmental Systems}, vol.~8, no.~2, pp. 128--138, 2016.

\bibitem{calvo2016perceptual}
M.~G. Calvo and L.~Nummenmaa, ``Perceptual and affective mechanisms in facial expression recognition: An integrative review,'' \emph{Cognition and Emotion}, vol.~30, no.~6, pp. 1081--1106, 2016.

\end{thebibliography}

\end{document}